\documentclass[12pt,aps,amsfonts,amsmath,prd,preprint,nofootinbib]{article}

\usepackage{setspace} 
\usepackage[pdftex]{graphicx}
\usepackage{dcolumn}
\usepackage{amssymb}
\usepackage{amsmath}
\usepackage[english]{babel}
\usepackage{bm}
\usepackage{cases}
\usepackage[pdftex]{hyperref}
\usepackage[svgnames]{xcolor}
\definecolor{phthaloblue}{rgb}{0.0, 0.06, 0.54}
\hypersetup{
    colorlinks=true,
    linkcolor=phthaloblue,
    citecolor=blue,
    filecolor=blue,
    urlcolor=phthaloblue,
    }
\makeatletter \def\@eqnnum{{\normalsize \normalcolor (\theequation)}}  \makeatother %

\begin{document}
\title{\bf Inflation in Random Landscapes \\
with two energy scales}

\author{Jose J. Blanco-Pillado$^{1,2}$\thanks{josejuan.blanco@ehu.es}, 
\ Alexander Vilenkin$^{3}$\thanks{vilenkin@cosmos.phy.tufts.edu}, 
\ Masaki Yamada$^{3}$\thanks{Masaki.Yamada@tufts.edu}\\[0.5cm]
$^{1}$ {\small IKERBASQUE, Basque Foundation for Science, 48011, Bilbao, Spain}\\
	$^{2}$ {\small Department of Theoretical Physics, University of the Basque Country,}  \\
	 {\small 48080, Bilbao, Spain}\\
	$^{3}$ {\small Institute of Cosmology, Department of Physics and Astronomy,} \\
	{\small Tufts University, Medford, MA  02155, USA}}

\def\({\left(}
\def\){\right)}
\def\[{\left[}
\def\]{\right]}
\def\det{{\rm det}}
\def\nn{\nonumber \\}
\def\p{\partial}
\def\pot{U}
\def\field{\phi}
\def\hs{\zeta}
\def\hess{\zeta}
\def\grad{\eta}
\def\thi{\rho}
\def\Tr{{\rm Tr}}
\def\Mpl{M_{\rm pl}}
\def\lmk{\left(}
\def\rmk{\right)}
\def\lkk{\left[}
\def\rkk{\right]}
\def\dd{{ d}}
\def\la{\left<}
\def\ra{\right>}
\newcommand{\eq}[1]{Eq.~(\ref{#1})}
\newcommand{\beq}{\begin{eqnarray}} 
\newcommand{\eeq}{\end{eqnarray}}
\newcommand{\bel}[1] {\begin{equation}\label{#1}}
\newcommand{\beal}[1] {\begin{eqnarray}\label{#1}}
\newcommand{\be}{\begin{equation}}
\newcommand{\ee}{\end{equation}}
\newcommand{\bea}{\begin{eqnarray}} 
\newcommand{\eea}{\end{eqnarray}}
\newcommand{\abs}[1]{\left\vert#1\right\vert}
\newcommand{\expec}[1]{\langle #1 \rangle}

\def\del{\partial}

\date{}
\maketitle
\onehalfspacing
\begin{abstract}
  We investigate inflation in a multi-dimensional landscape with a hierarchy of energy scales, motivated by the string theory, where the energy scale of Kahler moduli is usually assumed to be much lower than that of complex structure moduli and dilaton field.  We argue that in such a landscape, the dynamics of slow-roll inflation is governed by the low-energy potential, while the initial condition for inflation are determined by tunneling through high-energy barriers. We then use the scale factor cutoff measure to  calculate the probability distribution for the number of inflationary e-folds and the amplitude of density fluctuations $Q$, assuming that the low-energy landscape is described by a random Gaussian potential with a correlation length much smaller than $M_{\rm pl}$. We find that the distribution for $Q$ has a unique shape and a preferred domain, which depends on the parameters of the low-energy landscape.  We discuss some observational implications of this distribution and the constraints it imposes on the landscape parameters.

\end{abstract}
\flushbottom

\clearpage

\section{Introduction}
\label{sec:intro}

String theory is currently our best candidate for a fundamental theory of nature,
but its internal consistency requires it to live on a higher dimensional spacetime.  
This forces us to think of a mechanism of compactification that allows the theory
to be compatible with low energy observations. The effective four dimensional 
theory that results from this compactification process is then endowed with a large 
number of fields that parametrize the geometric properties of the
internal space. We therefore expect a large number of metastable vacua of the 
compactification potential where these
four dimensional fields would be stabilized.

Models of flux compatification, where 
fluxes are wrapped around the internal cycles of the compact manifold,
have been extensively studied in the literature \cite{Douglas:2006es}. This type of scenarios
leads to an incredibly large ensemble of vacua, due to the
huge numbers of combinatoric possibilities \cite{Bousso:2000xa}.

In order to explore the implications of the String Theory Landscape in cosmology, one needs to 
understand the basic properties of the compactification potential. Due to its intrinsic
complexity and the large number of fields involved, it seems reasonable to use statistical
techniques. Following these ideas it has been recently 
suggested  (see, e.g., \cite{Tegmark:2004qd,Aazami:2005jf,Frazer:2011tg,Battefeld:2012qx,Yang:2012jf,Bachlechner:2014rqa,Wang:2015rel,
Masoumi:2016eqo,Easther:2016ire,Masoumi:2016eag,Masoumi:2017gmh,Masoumi:2017xbe,Bjorkmo:2017nzd} )
\footnote{An alternative approach, based on the Dyson Brownian motion model, 
was developed in Refs.~\cite{Marsh:2013qca,Dias:2016slx,Wang:2016kzp,Freivogel:2016kxc,Pedro:2016sli,Dias:2017gva}.
}
that one could model the Landscape potential as a 
Gaussian Random Field (GRF) on the space of the low energy scalar fields, the moduli. This 
is a simple model designed to capture the random behavior of a
low-energy  potential that has a large number
of contributions of different physical origin.

Alternatively, one could take the opposite approach where one
investigates a particular set of potential realizations that appear in the
low energy description of a specific compactification scenario. This approach has also been
pursued in the literature for a small number of moduli fields in~\cite{BlancoPillado:2012cb, M, MartinezPedrera:2012rs}.

This kind of top-down approach, where one identifies a concrete
mechanisms that should be invoked to stabilize the moduli, highlights
the fact that not all the moduli should be treated in the same way.
In fact, it is pretty clear that all the models of compactification 
so far used in the literature work by introducing several ingredients
that stabilize some sectors of the moduli space but not others. This 
leads to the possibility of a hierarchy of scales in the compactification
process. One can take, for example, models of Type IIB compactification
to realize that the mechanisms that stabilize the complex structure moduli and
the dilaton are very different in nature than the one that fixes the Kahler moduli,
leading to a hierachy of masses~\cite{Conlon:2005ki,Gallego:2008qi,M}.

Following these considerations, in the present paper we investigate a Random 
Gaussian Landscape with a hierarchy of two different scales: a high-energy and 
low-energy sectors of the Landscape. 
We will show that this structure of the Landscape has important cosmological
implications. In particular, we will argue that transitions by bubble nucleation 
between the vacua in the low-energy landscape are likely to be subdominant. 
This implies that the initial conditions for slow-roll inflation, 
which occurs after tunneling from a false vacuum to a slow-roll region, 
are likely to be determined by tunneling through the barriers of the high-energy landscape.  
However, after the bubble nucleation, 
the dynamics of slow-roll inflation is governed by the low-energy landscape.

Thus, in a two-scale landscape, bubble nucleation 
and slow-roll inflation occur at different energy scales.\footnote{A similar suggestion
was used in the context of Type IIB compactifications in \cite{BlancoPillado:2012cb} where the complex structure
moduli were treated as a high energy random sector of the Landscape, while the Kahler 
moduli were considered a low energy sector.}
We use this fact to calculate 
the probability distribution for the maximal number of e-folds, $N_{\rm max}$, and the amplitude of 
density fluctuations $Q$ in the multiverse. We show that 
the probability function of $Q$ has a unique form and a preferred domain depending on parameters of the low energy landscape.

This paper is organized as follows. In the next section we specify the model of a two-scale landscape 
and argue 
{\it (i)} that tunnelings in the low-energy sector can be neglected and {\it (ii)} that the tunneling points that determine the initial conditions in the bubbles are randomly distributed in the low-energy landscape.
We then review some relevant features of slow-roll inflation.  In Sec.~\ref{sec:probability}, 
we calculate the prior probability distribution for the number of inflationary e-foldings and the amplitude of density fluctuations in a one-dimensional landscape.  (Here the word "prior" means "disregarding anthropic considerations".)
In Sec.~\ref{sec:multifield}, we extend this calculation to a multi-dimensional landscape and show that the resulting distribution is essentially the same as in one dimension.  The anthropic factor and some observational implications 
of our results are discussed in Sec.~\ref{sec:observation}.
Finally, our conclusions are summarized and discussed in Sec.~\ref{sec:conclusion}. 
We use the reduced Planck units ($M_{\rm pl} \simeq 2.4 \times 10^{18} \ {\rm GeV} \equiv 1$) throughout the paper.

\section{A two-scale Landscape}
\label{sec:two-scale}

We consider a model with two types of fields denoted as $\phi_i$ ($i = 1,2, \dots, D_{\rm L}$) and 
$\psi_j$ ($j = 1,2,\dots, D_{\rm H}$) that represent different sectors of the moduli space. In particular, one 
could think of $\phi_i$ as the Kahler moduli sector and $\psi_j$ as the complex structure moduli and the dilaton of a type IIB compactification.

 We assume that the potential $U({\boldsymbol \field}, {\bm \psi})$ 
has two characteristic energy scales $U_{\rm L}$ and $U_{\rm H}$ and two correlation lengths 
$\Lambda_{\rm L}$ and $\Lambda_{\rm H}$ in the field space, where 
the correlation function of the potential 
$F (|\boldsymbol \field_1 - \boldsymbol \field_2|, |\boldsymbol \psi_1 - \boldsymbol \psi_2|)$ 
rapidly decays at $|\boldsymbol \field_1 - \boldsymbol \field_2| \gg \Lambda_{\rm L}$ 
or $|\boldsymbol \psi_1 - \boldsymbol \psi_2 | \gg \Lambda_{\rm H}$.
We assume also that the potential changes by $\sim U_{\rm L}$ 
when the field value changes by $\Delta \boldsymbol \field \sim \Lambda_{\rm L}$, 
while it changes by $\sim U_{\rm H}$ when $\Delta \boldsymbol \psi \sim \Lambda_{\rm H}$. 
We consider the case where $U_{\rm L} \ll U_{\rm H}$ and $\Lambda_{\rm L}, \Lambda_{\rm H} \ll 1$. 
We shall also assume that the two correlation lengths are not much different from one another,%
\footnote{
This assumption is not essential for our analysis; it would be sufficient to consider the case where 
$(\Lambda_L/\Lambda_H)^4 (U_H/U_L) \gg 1$.
} 
\beq
\Lambda_L \sim \Lambda_H.  
\label{LLLH}
\eeq

According to the effective theory perspective, 
we expect that the potential of scalar fields is
well described by a Taylor expansion in the neighborhood of any point in the field space. 
Since the potential is characterized by $U_a$ and $\Lambda_a$ ($a = {\rm L}, {\rm H}$), 
we expect that the typical values of Taylor coefficients are given by 
\beq
 \frac{\del^n U}{\del \phi_i^n} \sim \frac{U_{\rm L}}{\Lambda_{\rm L}^n} 
 \qquad 
  \frac{\del^n U}{\del \psi_i^n} \sim \frac{U_{\rm H}}{\Lambda_{\rm H}^n}, 
 \qquad 
  \frac{\del^{m+n} U}{\del \phi_i^m \del \psi_j^n} \sim \frac{U_{\rm L}}{\Lambda_{\rm L}^m \Lambda_{\rm H}^n}, 
  \label{typical values}
\eeq
at a generic point in the landscape. The probability distribution of Taylor coefficients 
can be derived once we specify the correlation function $F (|\boldsymbol \field_1 - \boldsymbol \field_2|, | {\bm \psi}_1 - {\bm \psi}_2|)$.

Here we explicitly write an example of such a correlation function. 
Let us first decompose the potential into two terms plus a constant term $\bar{U}_{T}$
\beq
 U ( {\bm \phi}, {\bm \psi}) = U_{\rm H} ({\bm \psi}) + U_{\rm L} ( {\bm \phi}, {\bm \psi}) + \bar{U}_{T}. 
 \label{UHUL}
\eeq
Then, in order to satisfy the properties mentioned earlier we can take the functions $U_{\rm H} ({\bm \psi})$ and $U_{\rm L} ( {\bm \phi}, {\bm \psi})$
as Gaussian Random Fields with the following properties
\beq
 &&\la U_{\rm H} ({\bm \psi})\ra 
 = \la U_{\rm L} ({\bm \phi}, {\bm \psi})\ra =0 
 \label{1}
 \\
 &&\la U_{\rm H} ({\bm \psi}_1) U_{\rm H}({\bm \psi}_2) \ra 
 = U_{\rm H}^2 F_{\rm H} \lmk \frac{\Delta \psi}{\Lambda_{\rm H}} \rmk 
 \nonumber
 \\
 &&\la U_{\rm L} ({\bm \phi}_1, {\bm \psi}_1) U_{\rm L}({\bm \phi}_2, {\bm \psi}_2) \ra 
 = U_{\rm L}^2 
 F_{\rm L} \lmk \frac{\Delta \phi}{\Lambda_{\rm L}}, \frac{\Delta \psi}{\Lambda_{\rm H}} \rmk  
 \nonumber
 \\
 &&\la U_{\rm H} ({\bm \psi}_1) U_{\rm L}({\bm \phi}_2, {\bm \psi}_2) \ra 
 = 0, 
\label{2}
\eeq
where $\Delta \psi \equiv | {\bm \psi}_1 - {\bm \psi}_2|$ 
and $\Delta \phi \equiv | {\bm \phi}_1 - {\bm \phi}_2|$.  The functions $F_{\rm H} (x)$ and $F_{\rm L}(x,y)$ decay rapidly at $x \gg 1$ and/or $y \gg 1$. 
The correlators are often chosen in the form $F_{\rm H} (x) \propto \exp(-x^2 /2)$ and $F_{\rm L}(x,y) \propto \exp ( - x^2 /2 - y^2/ 2)$.  
However, this is a very special case, in which the statistics of the potential minima is rather different from that for a generic correlator. 
In particular, the minima are much stronger localized in energy in the limit of large $D$~\cite{Bray:2007tf}.  In this paper 
we focus on the generic case, as it was done in Ref.~\cite{Masoumi:2017xbe}. 

The minima of the Gaussian random field $U_{\rm H} ({\bm \psi})$ centered around zero and a characteristic
scale of $ U_{\rm H}$ like the one we use here for the high-energy sector of our Landscape are localized 
at $U\sim -\sqrt{D_{\rm H}} U_{\rm H}$ 
within a range of $\Delta U\sim U_{\rm H}$~\cite{Bray:2007tf}.  We shall therefore assume that the constant term 
$\bar{U}_T$ is of the order $\sqrt{D_{\rm H}} U_{\rm H}$, so that we do not have to worry that almost all minima 
of $U ( {\bm \phi}, {\bm \psi})$ are at $U < 0$.  Alternatively, one might add a term like $m^2 {\bm \psi}^2$ with 
$m^2 \ll U_{\rm H} / \Lambda_{\rm H}^2$, instead of a constant $\bar{U}_{T}$, 
so that $\la U({\bm \phi}, {\bm \psi}) \ra$ can be as large as $\sqrt{D_{\rm H}} U_{\rm H}$ somewhere in the landscape.  
Such a term could arise due to mixing between axions and flux fields \cite{Dvali:2001sm, Dvali:2004tma}. 

Random potentials specified by the correlators (\ref{1}) and (\ref{2}) reflect the hierarchy between different sectors in the String Theory Landscape, although it is likely that the specific forms of the potentials that can be computed will not fall under this simple statistical
description. On the other hand, we believe that our results should have a wider applicability than the model (\ref{1}) and (\ref{2}).  The key assumption that we are going to adopt is that the low-energy potential $U_{\rm L} ( {\bm \phi}, {\bm \psi})$ at a fixed $\psi$ is a random Gaussian field.  However, our conclusions are rather insensitive to the assumptions we make about the high-energy landscape $U_{\rm H} ({\bm \psi})$, as long as a few basic conditions are satisfied.  These conditions are listed in section \ref{sec:conclusion} and are likely to hold in a wide class of models.  

Of course the Gaussian nature of the low-energy potential
is a strong assumption (see for example \cite{M,Brodie:2015kza,Marsh:2015zoa}), so such models should be regarded only as toy models for the string theory landscape.  
A realistic landscape is likely to include some runaway directions in which some of the compact dimensions decompactify~\cite{Dine:1985he} (see also, e.g., Ref.~\cite{Bachlechner:2016mtp}).  Such runaway potentials do not occur in a random Gaussian field.  Moreover, destabilization of the volume modulus, resulting in a complete decompactification, should lead to a state with $U=0$, which is inconsistent with the separation into high and low-energy sectors (since most of the vacua in the landscape have $U\sim U_H$).\footnote{We thank the anonymous referee for pointing out this issue. }
On the other hand, 
it may be an adequate approximation to
focus on the central region of the moduli space containing
stable vacua. The energy scale of Kahler moduli in that
region may be much smaller than that of complex structure
moduli, since the stabilization mechanisms are different
for these fields. Our two-scale model may thus be a useful
toy model for the relevant part of the landscape, which includes metastable vacua (and excludes the unstable runaway configurations). 
It would be interesting to investigate the issues discussed in this paper in a more realistic setup where ensembles of potentials can be obtained.

\subsection{Tunneling transitions}
\label{sec:volume fraction}

In the cosmological context, each positive-energy vacuum becomes a site of eternal inflation, and transitions between different vacua constantly occur via bubble nucleation, resulting in a multiverse of bubbles with diverse properties.  The initial conditions in each bubble are determined by the instanton describing nucleation of that bubble from its parent vacuum.
The aim of this paper is to calculate the probability distribution for some observables in this multiverse. 

Slow-roll inflation must have occurred in our region after the tunneling event that led to the formation of our bubble.  Since we cannot observe anything prior to the tunneling, 
all observables can be calculated once we specify the tunneling endpoint in the landscape (which is also the initial point of the slow-roll inflation).

Observational predictions in multiverse models depend on one's choice of the probability measure.  
Different measure prescriptions can give vastly different answers.  (For a review of this ``measure problem" 
see, e.g.,~\cite{Freivogel:2011eg}.)  However, measures that are free from obvious pathologies, such as the 
scale factor~\cite{Linde:1993nz, Linde:1993xx, DeSimone:2008bq,Bousso:2008hz}, lightcone~\cite{Bousso:2008hz} and watcher~\cite{Garriga:2012bc, Vilenkin:2013loa} measures, tend to make similar predictions.  For definiteness we shall use the scale-factor cutoff measure.  The probability of observing a certain kind 
of region is then proportional to the average number of observers in such regions under the cutoff 
surface of a constant scale factor $a$ in the limit of $a\to\infty$.

Let us label the vacua in the landscape by an index $i$. 
The probability of observing a bubble of type $i$ nucleated in a false vacuum $j$ 
can then be roughly estimated as (see, e.g., Ref.~\cite{DeSimone:2008if})
\beq
 P_{ij} \propto n_{ij}^{(\rm obs)} \kappa_{ij} s_j~, 
\label{Pij}
\eeq
where $n_{ij}^{(\rm obs)}$ is the number of observers per unit mass in this type of bubble.
We focus on bubbles having the same low-energy microphysics as ours, described 
by the Standard Model.  Then $n_{ij}^{(\rm obs)}$ depends on $Q$, on the present cosmological constant 
$\rho_v$, and on the curvature parameter $\Omega_c$, which is determined by the number of e-folds of 
slow-roll inflation.

$\kappa_{ij}$  in Eq.~(\ref{Pij}) is the transition rate from $j$ to $i$ per Hubble volume per Hubble time, 
and $s_j$ is the fraction of inflating volume in parent vacuum $j$ on a constant scale factor slice.
$s_j$ is proportional to the $j$-th component of the eigenvector of the following matrix 
with the largest eigenvalue~\cite{Garriga:1997ef, Garriga:2005av}: 
\beq
 M_{ij} = \kappa_{ij} - \delta_{ij} \sum_r \kappa_{r i}~. 
\eeq

Since there are two energy scales in the landscape, there are two types of vacuum transitions, corresponding 
to tunnelings through high- and low-energy barriers.  The corresponding transition rates can be vastly 
different. They can be estimated as $\kappa_{ij}\propto \exp(-S_{ij})$, where $S_{ij}$ is the instanton action and can be written as
\beq
S_{ij}=\frac{\Lambda_a^4}{U_a} {\bar{S}}_{ij} ~, 
\label{Sij}
\eeq
by a rescale of variables. 
Here, $a = {\rm L}$ or ${\rm H}$ and ${\bar{S}}_{ij}$ is independent of $U_a$ and $\Lambda_a$. 
In a one-dimensional Gaussian landscape, the rescaled action ${\bar{S}}_{ij}$ typically takes values between $\sim 10$ and $\sim 10^4$ \cite{Masoumi:2017gmh}.  One can expect a similar range for ${\bar{S}}_{ij}$ in a higher-dimensional landscape.

With $U_H\gg U_L$ and $\Lambda_H\sim \Lambda_L$, Eq.~(\ref{Sij}) tells 
us that vacuum transitions in the low-energy sector are very strongly suppressed.  The probability of 
observing anthropic bubbles resulting from such transitions is then negligibly small; hence we shall 
concentrate on bubbles produced by tunnelings through high-energy barriers.

To a good approximation, an instanton describing such a tunneling can be found by solving the Euclidean 
equations of motion for the fields ${\bm \psi}$ in the potential  $U_H({\bm\psi})$, disregarding their 
interaction with the low-energy fields ${\bm\phi}$.  The typical size of the instanton 
(i.e., the initial bubble radius) is then 
\beq
r_0\sim \frac{\Lambda_{\rm H}}{ \sqrt{U_{\rm H}}}.
\label{r0}
\eeq

The tunneling endpoint is typically very close to a local minimum of $U_{\rm H} ({\bm \psi})$ 
for a generic potential~\cite{Garriga:2013cix}, 
but we do not assume this in the following analysis.

We next consider the Euclidean equations of motion for the ${\bm \phi}$ fields, 
\beq
 \frac{\dd^2 \phi_i}{\dd r^2} + \frac{3}{r} \frac{\dd \phi_i}{\dd r} - 
 \frac{\del U_{\rm L}}{\del \phi_i} = 0~. 
 \label{separateEoM2}
\eeq

The displacement of ${\bm \phi}$ due to the tunneling process can be estimated as
\beq
 \Delta {\bm \phi} 
 &\sim& \frac{U_{\rm L}}{\Lambda_{\rm L}} r_0^2 
 \\
 &\sim& \Lambda_{\rm L} \frac{\Lambda_{\rm H}^2}{\Lambda_{\rm L}^2} \frac{U_{\rm L}}{U_{\rm H}} \ll \Lambda_L, 
\eeq
where we have used $\del U_{\rm L} / \del \phi_i \sim U_{\rm L} / \Lambda_{\rm L}$ and Eqs.~(\ref{LLLH}) 
and (\ref{r0}).  This indicates that the low-energy fields ${\bm \phi}$ remain largely unperturbed during a 
high-energy tunneling.

To check this qualitative argument, we consider the following toy model with only two fields, $\phi$ and $\psi$: 
\beq
 &&U(\phi, \psi)  = U_{\rm H} \lkk \frac{c_2}{2} \gamma \lmk \frac{\phi}{\Lambda_{\rm L}} \rmk^2 
 + \frac{c_3}{6} \gamma \lmk \frac{\phi}{\Lambda_{\rm L}} \rmk^3 
 + \frac{c_\psi}{2} \lmk \frac{ \psi}{\Lambda_{\rm H}} \rmk^2 \rkk \times \nonumber \\
 &&~~~~~~~~~~~~~~\times  \lkk \frac{c_{\phi f}}{2} \gamma  \lmk \frac{\phi}{\Lambda_{\rm L}} - R \cos \theta \rmk^2 + \frac{c_{\psi f}}{2} \lmk \frac{\psi}{\Lambda_{\rm H}} - R \sin \theta \rmk^2 
 + c_h \rkk 
 \label{toy model}
\eeq
where 
$ \gamma \equiv U_{\rm L} / U_{\rm H} \ll 1$. 
The parameters $c_2$, $c_3$, $c_\psi$, $c_{1f}$, $c_{2f}$, $R$, $\theta$, $c_h$ are assumed to be $O(1)$; 
we take $c_i = 1$ ($i = 3,\psi, \phi f, \psi f$), $c_2 = 0$, $c_h = 0.5$ and $R = 5$ as an example.  The parameters $R$ 
and $\theta$ determine the location of the false vacuum, and $c_h$ determines its energy density.  
An example of this potential  with $\theta = 2 \pi / 5$, $\Lambda_L=\Lambda_H$ and $\gamma=U_{\rm L}/U_{\rm H} = 0.02$
is shown in Fig.~\ref{fig:potential}, 
where the false vacuum is marked by a blue dot. 
We also show in Fig.~\ref{fig:potential} an inflection point with a white 
dot. In the following section we will consider this kind of region as one of the forms of the
potential to be responsible for the inflationary period after the tunneling event.

\begin{figure}[t] 
   \centering
   \includegraphics[width=5in]{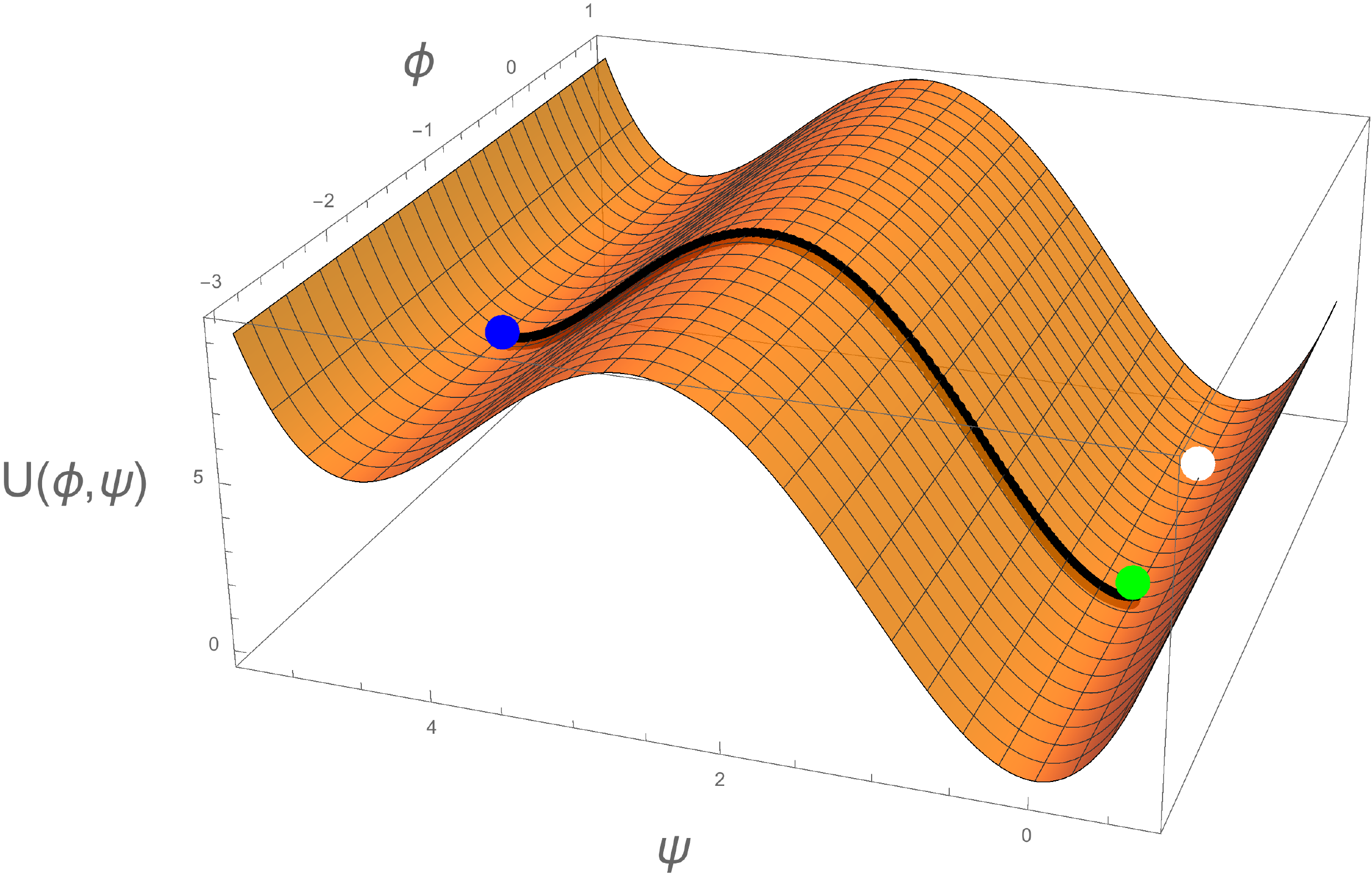} 
   \caption{
An example of the potential (\ref{toy model}), where we use the parameters indicated in the text. 
The false vacuum is marked by a blue dot. 
The black line shows the instanton trajectory, and the green dot is the tunneling point. 
We also show an inflection point with a white dot. 
   }
   \label{fig:potential}
\end{figure}

We used the public code developed in Ref.~\cite{Masoumi:2016wot} to solve the Euclidean equations of motion 
and determine the tunneling (end)point $(\phi, \psi) = (\phi_0, \psi_0)$ corresponding to bubble nucleation. 
We assume that gravitational effects on the tunneling can be neglected, which is usually the case for $\Lambda_a \ll 1$.  
The instanton trajectory is shown as a black line in the figure, with the tunneling point indicated by a green dot.

To illustrate the dependence of the tunneling point on $\gamma \equiv U_{\rm L} / U_{\rm H}$, 
we plot $\abs{\phi_0 - \phi_{\rm FV}}/|\phi_{\rm FV}|$ as a function of $\gamma$ in Fig.~\ref{fig:initial}, 
where $\phi_{\rm FV}$ is the field value at the false vacuum. 
We see that for $\gamma\ll 1$ the field $\phi$ changes very little due to the tunneling process, as expected.

\begin{figure}[t] 
   \centering
   \includegraphics[width=4.5in]{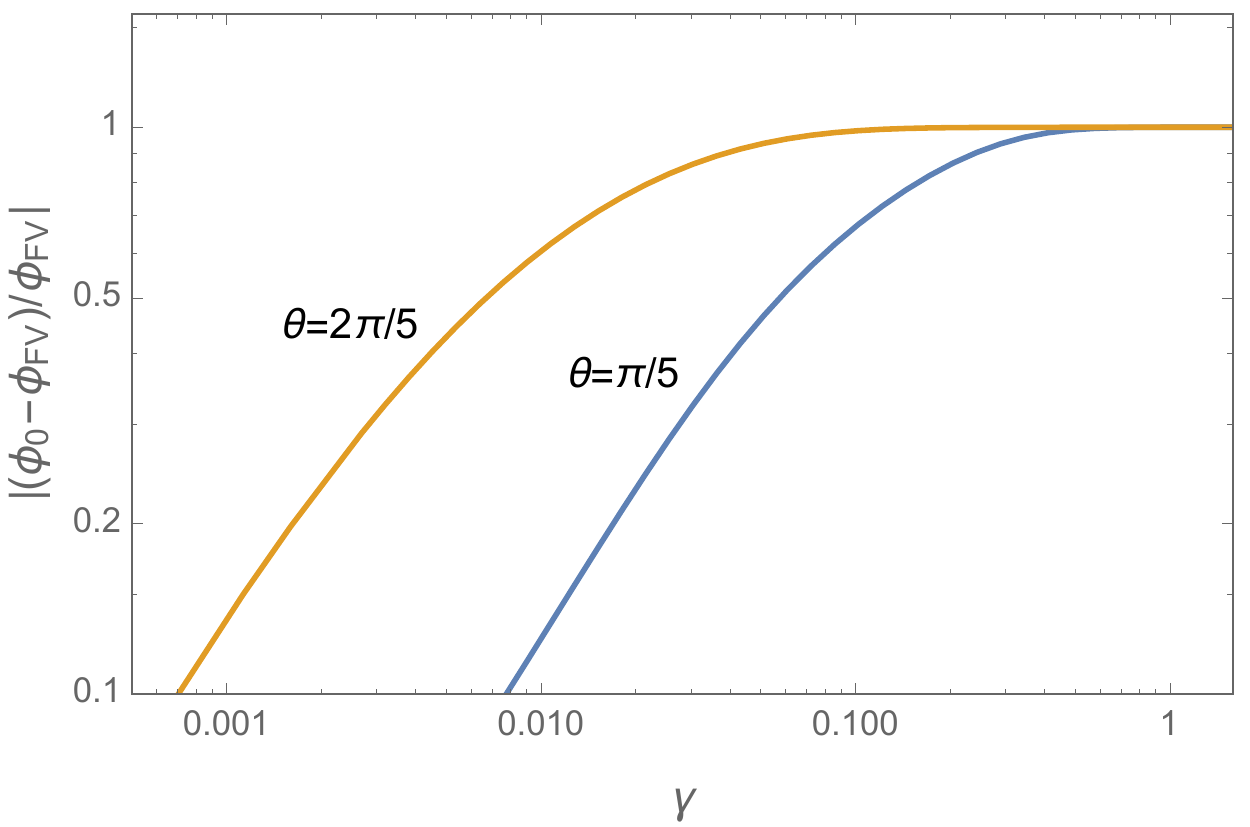} 
   \caption{
 Shift of the tunneling point ($\abs{(\phi_0 - \phi_{\rm FV})/\phi_{\rm FV}}$) as a function 
of $\gamma$ ($\equiv U_{\rm L} / U_{\rm H}$) for $\theta = \pi / 5$ (blue line) and $2 \pi / 5$ (orange line).  
   }
   \label{fig:initial}
\end{figure}

Since the high- and low-energy potentials in the landscape are assumed to be uncorrelated, we expect the distribution of tunneling points in the ${\bm\phi}$-space to be random -- that is, 
uncorrelated with the potential $U_{\rm L}({\bm \phi},{\bm \psi})$.

Finally, note that although we considered the case where $U_{\rm H} \gg U_{\rm L}$ 
and $\Lambda_{\rm H} = \Lambda_{\rm L}$ in the toy model, we expect
the same conclusion to apply in the case where $U_{\rm H} = U_{\rm L}$ and $\Lambda_{\rm H} \ll \Lambda_{\rm L}$. 
The latter case is somewhat similar to the model proposed in Ref.~\cite{Linde:2016uec}, 
where the dynamical scale of the inflaton $(\Lambda_{\rm L})$ 
is effectively stretched to infinity due to a singularity of the kinetic term.

\subsection{Slow roll inflation}

We now consider the dynamics of scalar fields after the tunneling.  As we mentioned earlier, 
the exit point of the tunneling process for the ${\bm \psi}$ field tends to be close to 
a local minimum of $U_H({\bm\psi})$, so in these cases, one can neglect its dynamics after the tunneling. 

Furthermore, we will now show that even if the tunneling point is not very close to its minimum, the subsequent 
evolution of the fields would not be much affected by this fact.  In order to do this we first note that, as is well known, the initial evolution
of the bubble is described by an open universe dominated by curvature. Given the hierarchy
of masses between the $\psi_i$ and $\phi_j$ fields, we expect the field $\psi$ to start
rolling first towards its minimum at a time $t_H \sim m_H^{-1}$, where $m_H \sim U_H^{1/2} / \Lambda_H$ is the typical
mass for the moduli in the high energy sector. The energy of this
oscillating field around its minimum would redshift with the expansion of the universe inside the bubble as
a matter energy component, $\rho_{\psi} \sim a^{-3}$, so it will continue to be subdominant 
compared to the curvature term $\rho_K \sim a^{-2}$. This will remain to be the case until the inflationary energy density becomes 
relevant at a much later time of the order $t_L \sim U_L^{-1/2}$. By then the amplitude of
the $\psi$ field would be suppressed by the expansion of the universe during this time and 
one can consider it at its minimum.

One can also study the evolution of the $\phi$ field during this time.  Taking 
into account the low mass of this field, one should consider the possible coupling between the $\phi$ and $\psi$ fields
as the dominant effect. We will denote this interaction by a term in the Lagrangian of the form $\sim \rho_{\psi\psi\phi} \psi^2 \phi$.\footnote{ We use the notation $\rho_{\psi\psi\phi}$ for the coupling constant for consistency with the notation in Ref.~\cite{Masoumi:2017xbe}.  It should not be confused with the energy density $\rho_\psi$.} This evolution was studied in 
detail in  Ref.~\cite{Masoumi:2017xbe} where it was found that the effect of this interaction in a 
background of an oscillating $\psi$ field would be to
shift the value of $\phi$ by a constant of the order
\beq
\Delta \phi_{\text{curv}} \sim {{\rho_{\psi\psi\phi} \psi_0^2}\over{m_H^2}}
\eeq
where $\psi_0$ is the initial deviation of $\psi$ with respect to its minimum, its initial amplitude. Using
generic values of these coefficients in our landscape we arrive at,
\beq
\Delta\phi_{\text{curv}} \sim \left( \frac{U_L}{U_H} \right) \left(\frac {\Lambda_H^2}{\Lambda_L^2} \right) \left(\frac{\psi_0^2}{\Lambda_H^2} \right) \Lambda_L \ll \Lambda_L
\eeq
where we have used that $\rho_{\psi\psi\phi}\sim U_L/\Lambda_H^2\Lambda_L$, $\abs{{\bm \psi_0}} \lesssim \Lambda_{\rm H}$ and $U_{\rm L} \Lambda_{\rm H}^2 / ( U_{\rm H} \Lambda_{\rm L}^2) \ll 1$.
This implies that we can indeed neglect the evolution of the fields
after the tunneling even if the exit point is not very close to the minimum of the potential.

Thus the relevant dynamics after $t \sim t_L$ is that of ${\bm \phi}$ whose potential is  
\beq
U ({\bm \phi}) = U_{\rm L} ({\bm \phi}, {\bm \psi}_0) + \bar{U} 
\label{Uphi}
\eeq
with ${\bm \psi}_0$ fixed and where we have introduced the quantity
\beq
\bar{U} \equiv  U_{\rm H} ({\bm \psi}_0) + \bar{U}_{T}~.
\label{barU}
\eeq
The correlator of this potential is given by 
\beq
 \la U ({\bm \phi}_1) U ({\bm \phi}_2) \ra - \bar{U}^2 
 = U_L^2 
 F_{\rm L} \lmk \frac{\Delta \phi}{\Lambda_L},0 \rmk. 
\eeq

The magnitude of $\bar{U}$ is different for different tunnelings, with a typical range of variation $\sim U_H$.  However, 
the anthropic argument \cite{Weinberg:1987dv, Weinberg:1988cp} requires that 
the cosmological constant after slow-roll inflation should (almost) vanish, 
and thus we are interested only in tunnelings for which the two terms in (\ref{barU}) nearly cancel, so that
$\bar{U}\sim U_{\rm L}$.  This will be discussed later in detail. 
In the rest of this paper, we 
denote $U_{\rm L}$ as $U_0$, $\Lambda_{\rm L}$ as $\Lambda$, and $D_{\rm L}$ as $D$ 
for notational simplicity.

We focus on the case of small-field inflation where $\Lambda \ll 1$. 
Then $(U'/U)^2$, $U'' / U \sim 1/\Lambda^2 \gg 1$ at a generic point, so 
slow-roll inflation can occur only in rare regions.  It typically occurs either near inflection 
or saddle points, where the required fine-tuning of Taylor coefficients is minimal \cite{Linde:2007jn,Baumann:2007ah}.  
At such points the potential is accidentally flat in one of the field directions, while its curvature 
is expected to be large in all other directions.  We denote the flat direction as $\phi$ and the 
other directions, which are perpendicular to $\phi$, as $\phi_\perp$. 

It was shown in \cite{Masoumi:2017xbe} that small-field inflation in a random Gaussian landscape is typically 
single-field, with the fields $\phi_\perp$ playing no dynamical role.   
Let us then briefly review some properties of inflection and saddle point inflation~\cite{Linde:2007jn,Baumann:2007ah}, 
neglecting the perpendicular directions. 
The potential is written as 
\beq
U(\phi)=U + \eta\phi + \frac{1}{2} \lambda\phi^2 + \frac{1}{3!}\rho \phi^3 +..., 
\label{Uphi}
\eeq
where we assume $\lambda=0$ with $\eta\rho >0$ for inflection-point inflation and $\eta=0$ 
with $\lambda<0$ for saddle point inflation \footnote{Note that $\eta$ parametrizes the first derivative 
of the potential and it is not to be confused with the slow roll parameter.}. We define the slow-roll range 
as the region of field space that satisfies the slow-roll condition, $|U''/U| \le 1$, or 
\beq
 \abs{\phi} \lesssim \frac{U}{ \rho}~. 
 \label{SRrange}
\eeq

The maximal number of e-folds for inflection point inflation is given by 
\beq
N_{\rm max} \approx \int_{-\phi_{\rm end}}^{\phi_{\rm end}} d\phi \frac{U(\phi)}{U'(\phi)} \approx \pi\sqrt{2}\frac{U}{\sqrt{\eta\rho}}.
\label{Nmax}
\eeq
For a saddle point, eternal inflation occurs near the top of the potential, 
$|\phi|\lesssim U^{3/2}/|\lambda| \equiv \phi_q$, due to quantum diffusion~\cite{Vilenkin:1983xq}, 
so the integral in Eq.~(\ref{Nmax}) would diverge.  Nonetheless it will be useful to define%
\footnote{
The maximum number of e-folds in saddle point inflation could be defined as
\beq
 N_{\rm max} = \int_{\phi_q}^{\phi_{\rm end}} \frac{U}{U'} d U 
 \approx  (U/| \lambda|) \ln(|\lambda|/\rho U^{1/2})~. 
\eeq
This represents the number of e-folds in the slow-roll regime
after the diffusion region at the top of the potential.
It differs from (\ref{Nmaxsaddle}) only by a logarithmic factor. 
}
\beq
N_{\rm max} \equiv 2\pi\frac{U}{|\lambda|}~. 
\label{Nmaxsaddle}
\eeq

The magnitude of density fluctuations produced at observable scales is 
\beq
Q^2=\frac{1}{12\pi} \frac{U^3}{{U'}^2} 
\simeq \frac {N_{\rm CMB}^4\rho^2}{48 \pi U} f^2 \lmk 
x, y
\rmk
\label{Q^2}
\eeq
where $N_{\rm CMB}$ ($\approx 50-60$) is the e-folding number at which the CMB scale leaves the horizon
and we have introduced the dimensionless quantities
\beq
 x \equiv \pi \frac{N_{\rm CMB}}{N_{\rm max}}, 
 \label{x}
 \\
 y \equiv \frac{N_{\rm max}}{2 \pi}, 
 \label{y}
\eeq
which parametrize the function $f(x,y)$ defined in Appendix~\ref{sec:observables}.  This function is ${\cal O}(1)$ 
for ${N_{\rm max}}\gtrsim N_{\rm CMB}\gg 1$.

The actual number $N_e$ of slow-roll e-folds depends on the inititial conditions after tunneling.
If the tunneling point is outside the slow-roll range, the field starts rolling fast (after a brief 
curvature-dominated period) and may overshoot part or all of the slow-roll region. 
It was shown in Ref.~\cite{Masoumi:2017gmh} that if the tunneling point is in the range 
\beq
- v_0 \frac{U}{\rho} \lesssim \phi \lesssim -\frac{U}{\rho}~,
\label{range}
\eeq 
where $v_0 \simeq 17$ is a numerical constant, then the field slows down and undergoes a nearly 
maximal number of inflationary e-folds so, in this case, we have $N_e \approx N_{\rm max}$. On either side of this range, 
$N_e$ drops towards zero within a distance $\Delta \phi \sim U/\rho$~.\footnote{In the case of saddle point inflation, the attractor region consists of two segments 
separated by a large gap [$\phi \sim ( - v_0 U / \rho, 0)$], where the field ends up on the ``wrong'' side of the hill and rolls into the shallow minimum next to the saddle point. 
The attractor range of inflation is in the intervals $\Delta \phi \sim U / \rho$ 
near the boundaries of this range. 
Thus the size of the attractor region is approximately given by $U/\rho$ 
without the factor $v_0$. 
}

We now comment on the dynamics of the fields $\phi_\perp$ after the tunneling. These fields 
typically have large initial displacements, $\phi_\perp \sim\Lambda$, and the potential $U({\bm \phi})$ causes 
them to oscillate. But oscillations are quickly damped and the fields settle at some point on the $\phi$-axis,
where $\del U / \del \phi_\perp = 0$.  Slow-roll inflation occurs if $\phi$ at that point is in 
the range (\ref{range}).  The region of ${\bm\phi}$-space encompassing all tunneling points 
that lead to inflation about a given inflection or saddle point will be called the attractor region 
of that point.  It can be characterized by the volume fraction $ f_{\rm vol}$ that it occupies in a 
correlation volume $\Lambda^D$.
It has been shown in Ref.~\cite{Masoumi:2017xbe} that 
\beq
 f_{\rm vol} \sim \frac{v_0 U /\rho}{\Lambda}, 
\eeq
the same as in the one dimensional case.\footnote{This can be understood as follows.  While 
the $\phi_\perp$ fields oscillate, the dynamics of $\phi$ is driven mostly by the interaction term 
$\propto\phi_\perp^2 \phi$.  (This is because the gradient of the $\phi$-potential in (\ref{Uphi}) 
is very small.)  The oscillation time scale is short compared to the slow roll of $\phi$; hence we 
can average over the oscillations.  This gives a time-dependent force term for $\phi$, resulting 
in a shift of $\phi$ which is independent of its initial value (but does depend on the initial amplitude 
of $\phi_\perp$).  It follows that the width of the attractor region is the same as in $1D$ for all 
values of $\phi_\perp$.}

\section{Probability distribution for observables}
\label{sec:probability}

As we already mentioned, we expect the distribution of tunneling points in the ${\bm\phi}$-space to be random -- that is, 
uncorrelated with the potential $U({\bm \phi})$.  This is because the tunnelings are governed by the high-energy potential, which is uncorrelated with the low-energy potential.  For the same reason,
the tunneling rate $\kappa_{ij}$ and the parent 
vacuum volume fraction $s_j$ in Eq.~(\ref{Pij}) should also be uncorrelated with the location of 
the tunneling point in the low-energy landscape.  It then follows that the probability of observing 
a bubble that resulted from tunneling to a vicinity of an inflection or saddle point $a$ is given by 
\beq
 P_a  \propto  n_a^{(\rm obs)} f_{{\rm vol},a}, 
\eeq
where $f_{{\rm vol},a}$ ($\sim v_0 U / (\rho \Lambda)$) 
is the volume fraction of the corresponding attractor region.\footnote{Tunneling to a close 
vicinity of a saddle point may result in a quantum diffusion regime of eternal inflation, which 
gives rise to an unlimited number of anthropic ``pockets", each of which  ultimately produces 
an infinite number of observers.  Naively, one might think that this would make observing saddle 
point inflation infinitely more probable.   However, this is not the case.  In the scale factor 
measure, the counting of observers is dominated by the bubbles that are formed near the 
cutoff surface, so the infinite numbers of pockets and observers formed afterwards are irrelevant.}

We are interested in the probability distribution for some observables in the multiverse. 
The distribution of $N_{\rm max}$ is useful to find the probability distribution for the spectral index $n_s$, 
because $n_s$ is rigidly correlated with $N_{\rm max}$ (see \eq{n_s1} and \eq{n_s2} in Appendix A). 
Other important observables are the amplitude of density perturbation $Q$ and 
the energy density of the present vacuum (or the cosmological constant) $\rho_v$ . 
The actual e-folding number $N_e$ is also important for calculating the curvature of the present Universe. 
Since $N_e$ is typically close to 
the maximal e-folding number $N_{\rm max}$ for the case of inflection point inflation, 
we do not need to calculate its distribution separately from $N_{\rm max}$. 
For the case of saddle point inflation, the distribution of $N_e$ is not relevant for observational 
predictions, because the range of spectral index $n_s$ predicted by saddle point inflation is already ruled out.

As already mentioned, we are focusing on vacua having the same low-energy microphysics 
as ours (the Standard Model) and differ only in the high-energy sector (including the inflaton). 
The probability distribution for $N_{\rm max}$, $Q$ and $\rho_v$ in the multiverse is then
\beq
 P(N_{\rm max},Q,\rho_v) 
 &=& 
 \sum_{a} P_{a} \delta \lmk N_{\rm max} - N_{{\rm max},a} \rmk \delta \lmk Q - Q_{a} \rmk 
 \delta \lmk \rho_v - \rho_{v,a} \rmk 
 \nonumber
 \\
 &\propto& n^{(\rm obs)} (N_{\rm max}, Q,\rho_v) 
 \int d \bar{U} P_H (\bar{U}) P_{L} (N_{\rm max},Q, \rho_v,\bar{U}) 
\eeq
where $a$ labels different vacua and we have introduced $P_H(\bar{U})$ as
the distribution of the values of $\bar{U}=U_{\rm H} ({\bm \psi}) + \bar{U}_{T}$ at a randomly chosen 
minimum in the high-energy landscape and
$P_{L} (N_{\rm max},Q, \rho_v,\bar{U}) $ as the
probability distribution for a low energy landscape
with a mean potential $\bar{U}$ to have an inflationary
region characterized by $N_{\rm max}$ and $Q$ and leading to a minimum
with a cosmological constant $ \rho_v$.
The volume fraction $f_{{\rm vol},a}$ is included in the definition of $P_L$.

We further assume that we can factorize the distribution
related to the low energy sector in the following way
\beq
P_{L} (N_{\rm max},Q, \rho_v,\bar{U})  = P_{NQU}(N_{\rm max},Q, \bar{U}) P_{\rm cc} (\rho_v; \bar{U}) 
\eeq
where, $P_{NQU}(N_{\rm max},Q, \bar{U})$ is the probability for a randomly chosen 
inflection or saddle point to have a given value of ${\bar U}$ and to yield the specified values 
of $N_{\rm max}$ and $Q$ and $P_{\rm cc} (\rho_v; \bar{U})$ is the energy distribution of potential minima 
in the low-energy landscape. This factorization is justified because the separation between the 
potential minimum and the inflection/saddle point is comparable to the correlation length $\Lambda$.
Using this factorization we arrive at
\beq
 P(N_{\rm max},Q,\rho_v) 
 &\propto& n^{(\rm obs)} (N_{\rm max}, Q,\rho_v) 
 \int d \bar{U} P_H (\bar{U})  P_{\rm cc} (\rho_v; \bar{U}) P_{NQU}(N_{\rm max},Q, \bar{U})  
 \label{P}
 \eeq
 According to Ref.~\cite{Bray:2007tf}, $P_{\rm cc} (\rho_v; \bar{U})$ is given by 
\beq
 P_{\rm cc} (\rho_v; \bar{U}) 
 \sim \frac{1}{U_{\rm L}} \exp \lkk - \lmk d_1 \frac{\rho_v - \bar{U}}{U_{\rm L}} 
 + d_2 \sqrt{D_{\rm L}}  
\rmk^2 
- d_3 \lmk \frac{\rho_v - \bar{U}}{U_{\rm L}} \rmk^2 \rkk, 
\label{rho_v}
\eeq
where $d_i$ are ${\cal O}(1)$ constants.  (Note that $D_{\rm L} = D$, $U_{\rm L} = U_0$, 
and $\Lambda_{\rm L} = \Lambda$ in our current notation).  As expected, this distribution 
is nearly flat in the anthropic range $\rho_v \sim (-10^{-120}, 10^{-120})$, so we can approximate 
$P_{\rm cc} (\rho_v; \bar{U})\approx P_{\rm cc} (0; \bar{U})$.

The stochastic variable $\bar{U}=U_{\rm H} ({\bm \psi}) + \bar{U}_{T}$ is independent of the 
low-energy potential, and  the $P_H(\bar{U})$ distribution varies on a characteristic scale 
$\sim U_{\rm H}$.  On the other hand, the factor $P_{\rm cc} (0; \bar{U})$ in Eq.~(\ref{P}) effectively 
restricts the range of integration to ${\bar U}\sim\sqrt{D_{\rm L}}  U_{\rm L}$ with a width of order 
$U_{\rm L}$, enforcing the condition that the potential difference between the slow-roll region and 
the minimum should be $\lesssim U_{\rm L}$.  Since this energy scale is much smaller than 
$U_{\rm H}$, $P_H(\bar{U})$ is approximately constant in this domain of integration: 
$P_H (\bar{U})\approx P_H(0)$.  And since the normalization of the distribution (\ref{P}) is not fixed, 
we shall drop this constant in what follows. 

Putting all this together, we can rewrite Eq.~(\ref{P}) as
\beq
 P(N_{\rm max},Q,\rho_v) \propto n^{(\rm obs)}(N_{\rm max}, Q,\rho_v) P_{NQ} (N_{\rm max}, Q),
\label{PnP}
\eeq
where
\beq
 P_{NQ} (N_{\rm max}, Q) = \int d{\bar U} P_{\rm cc} (0; \bar{U})  P_{NQU} (N_{\rm max},Q, \bar{U}). 
\label{prob}
\eeq
We shall refer to $P_{NQ}$ and $n^{(\rm obs)}$ as the "prior" distribution and the anthropic factor, respectively.

We shall calculate the prior distribution $P_{NQ}$ in Sec.~\ref{sec:multifield}.  As a warmup exercise, 
in the next subsection we shall calculate this distribution for the case of inflection point inflation 
in a one-dimensional landscape.  This is especially useful, since we will find later on that the 
calculation in the higher-dimensional case reduces to that in one dimension.  This is not surprising, 
since it was shown in Ref.~\cite{Masoumi:2017xbe} that small-field inflation is essentially one-dimensional.  The 
anthropic factor $n^{(\rm obs)}$ will be discussed in Sec.~\ref{sec:observation}.

\subsection{Prior probability distribution in a one-dimensional Landscape}

We consider a one-dimensional random Gaussian landscape $U(\phi)$ with characteristic energy 
scale $U_0$ and correlation length $\Lambda$. The average value of the potential ${\bar U}$ is 
assumed to be fixed (so we do not need to integrate over ${\bar U}$).  The probability that 
inflection-point inflation with certain values of $N_{\rm max}$ and $Q$ will occur with the 
initial value of $\phi$ randomly chosen in the landscape can then be calculated along the 
lines of Ref.~\cite{Masoumi:2016eag},  
\bea
\label{PQNU1}
P_{NQ}^{\rm (I)} (N_{\rm max},Q) 
=\frac{L}{{\cal N}_{\rm I}} \int dU d\eta d\lambda d\rho ~{\cal P}(U,\eta,\lambda,\rho) ~
|\rho| \delta(\lambda) ~
f_{\rm vol} (U, \rho) \times \nonumber \\
\times ~ \delta\left( N_{\rm max}-\frac{\pi\sqrt{2} U}{\sqrt{\eta\rho}}\right) 
 \delta\left(Q- \frac {N_{\rm CMB}^2\rho}{4\sqrt{3\pi U}} f(x,y) \right)~.
\eea
Here, ${\cal N}_{\rm I}$ is the number of inflection points and $L$ is the size of the landscape, so
${{\cal N}_{\rm I}}/{L}  \sim {1}/{\Lambda}$ is the density of inflection points.  The integration 
variables $U,\eta,\lambda,\rho$ are the coefficients in the Taylor expansion of the potential 
(\ref{Uphi}). Their distribution is given by
\beq
&&{\cal P}(U,\eta,\lambda,\rho)=
 A_1 A_2 \exp \lkk 
 - {\cal Q}_1 - {\cal Q}_2 \rkk, 
 \\
&& A_1 =  \frac{\lmk c_1 c_4 - c_2^2/4 \rmk^{1/2}}{\pi} \frac{\Lambda^2}{U_0^2}, 
\qquad 
 A_2  =\frac{ \lmk c_5 c_8 - c_6^2/4 \rmk^{1/2 } }{\pi } \frac{\Lambda^4}{U_0^2}
\\
&& {\cal Q}_1 = 
c_1 \frac{1}{U_0^2} (U - \bar{U})^2 - c_2 \frac{\Lambda^2}{U_0^2} (U - \bar{U}) \lambda 
+c_4 \frac{\Lambda^4}{U_0^2} \lambda^2, 
\qquad {\cal Q}_2 = 
c_5 \frac{\Lambda^2}{U_0^2} \eta^2 +c_6 \frac{\Lambda^4}{U_0^2} \eta \rho 
+c_8 \frac{\Lambda^6}{U_0^2} \rho^2 ~, \nonumber
\eeq
where $c_i$ can be expressed in terms of the moments of the correlation function 
and are ${\cal O}(1)$.  We also used Eqs.~(\ref{Nmax}) and (\ref{Q^2}) for $N_{\rm max}$ and $Q$ 
and included the volume factor $f_{\rm vol}(U,\rho)$ to account for the fact that the attractor region 
of an inflection point has size $\sim f_{\rm vol}\Lambda$.

Integrating out the delta functions in Eq. (\ref{PQNU1}) and using $f_{\rm vol} \sim v_0 U / (\rho \Lambda)$, 
we find \footnote{
When $v_0 U/\rho_* \sim \Lambda$, the slow-roll range becomes $\sim \Lambda$.  For smaller values 
of $\rho$ it remains $\sim \Lambda$, since the fourth and higher derivatives of $U$ become important.
}
\beq
P_{NQ}^{\rm (I)}(N_{\rm max},Q) \sim 
 \frac{2v_0}{N_{\rm max}Q} \int dU~ U \eta_* \rho_* ~{\cal P}(U,\eta_*,0,\rho_*) 
\label{P0}
\eeq
where
\beq
\eta_*=\frac{\pi^{3/2}}{2\sqrt{3}}\frac{U^{3/2}N_{\rm CMB}^2}{QN_{\rm max}^2} f(x,y), 
~~~ \rho_*=4\sqrt{3\pi}\frac{QU^{1/2}}{N_{\rm CMB}^2 f(x,y)},
\label{*0}
\eeq
are the values selected by the delta functions.

To analyze the distribution (\ref{P0}), we first note that we should have $U\lesssim U_0$, since 
higher values of $U$ are exponentially suppressed.  The slow roll condition requires 
$\eta_* \ll U \lesssim U_0$, while the typical value of $\eta$ in the landscape is $\eta_0\sim U_0/\Lambda \gg U_0$.
Hence, $\eta_*\ll \eta_0$, so we can set $\eta_*\approx 0$ in the exponent of (\ref{P0}).  Then
\beq
 {\cal P}(U,\eta_*,0,\rho_*) \approx A_1 A_2 \exp\left(-\frac{(U-\bar{U})^2}{U_0^2} -\frac{\rho_*^2}{\rho_0^2}\right) ,
\eeq
where we have defined $\rho_0 = U_0/\Lambda^3$ and  set $c_i \simeq 1$. 
Substituting this in (\ref{P0}), using (\ref{*0}) and disregarding numerical factors, we have
\beq
P_{NQ}^{\rm (I)} (N_{\rm max},Q) \sim \frac{v_0 \Lambda^6}{U_0^4} \frac{1}{Q N_{\rm max}^3} 
 \int dU~ U^{3} 
 \exp\left(-\frac{(U-\bar{U})^2}{U_0^2} -  \frac{48\pi Q^2}{N_{\rm CMB}^4 \rho_0^2 f^2(x,y)} U\right). 
\label{PNQ}
\eeq
 
Note that 
the slow-roll range (\ref{SRrange}) must be less than $\Lambda$,  so we need to impose the condition 
${U}/{\rho_*} \lesssim \Lambda$. 
Together with this condition, \eq{PNQ} is the final formula for $P_{NQ}^{\rm (I)}$.

We shall now estimate the shape of this distribution by approximating the integral in \eq{PNQ} in 
different regimes.  
First, the condition ${U}/{\rho_*} \lesssim \Lambda$ gives 
\beq
U \lesssim 48 \pi \frac{Q^2 \Lambda^2}{N_{\rm CMB}^4 f^2(x,y)} \equiv U_0 \frac{Q^2}{Q_1^2}. 
 \label{condition}
\eeq
If $Q \lesssim Q_1$, the integral is effectively cut off by this condition 
and we have 
\beq
 P_{NQ}^{\rm (I)}  (N_{\rm max}, Q) \sim \frac{v_0 \Lambda^{14} Q^7}{ U_0^4 N_{\rm CMB}^{16} N_{\rm max}^3 f^8(x,y)}, 
 ~~~~\lmk Q \lesssim 
Q_1 \rmk
\label{PNQ<}
\eeq

If $Q \gtrsim Q_1$, the character of the distribution (\ref{PNQ}) depends on the magnitude of the ratio
\beq
{\cal R} = \frac{\rho_*^2(U_0)}{\rho_0^2} \sim \frac{48\pi Q^2 \Lambda^6}{N_{\rm CMB}^4 U_0 f^2(x,y)} 
\equiv \frac{Q^2}{Q_2^2}. 
\label{R}
\eeq
We note that 
\beq
 \frac{Q_1}{Q_2} = \Lambda^2 \ll 1 ~. 
\eeq
If ${\cal R}\ll 1$, the integral is effectively cut off at $U\sim U_0$ and
\beq
P_{NQ}^{\rm (I)}  (N_{\rm max},Q) \sim \frac{v_0 \Lambda^6}{N_{\rm max}^3 Q}, 
~~~~\lmk Q_1 \lesssim Q \lesssim Q_2 \rmk. 
\label{PNQ<>}
\eeq
On the other hand, if ${\cal R}\gg 1$, the integration is cut off by the second term in the exponent and we have
\beq
 P_{NQ}^{\rm (I)}  (N_{\rm max}, Q) \sim \frac{v_0 N_{\rm CMB}^{16} U_0^4 f^8(x,y)}{N_{\rm max}^3 Q^9 \Lambda^{18}}, 
~~~~\lmk Q_2 \lesssim Q \rmk. 
\label{PNQ>}
\eeq
In this case inflation occurs at $U\ll U_0$.

We find that the $N_{\rm max}$ dependence is $N_{\rm max}^{-3}$ 
with a small correction coming from $f(x,y)$.%
\footnote{
Apart from a small correction, the dependence $P\propto N_{\rm max}^{-3}$ was first derived in 
Ref.~\cite{Agarwal:2011wm} in a simple model. 
}
The probability of inflation with $N_e > N_{\rm CMB}$ can be found by integrating the distribution 
(\ref{PNQ}) over $Q$ and over $N_{\rm max}$ from $N_{\rm CMB}$ to $\infty$. 
The $Q$-integral can be written as 
\beq
 \int_{Q_1}^\infty \frac{dQ}{Q} \exp \lkk- c_8 \frac{U}{U_0} \lmk\frac{Q}{Q_2}\rmk^2 \rkk.
\eeq
The integration is effectively cut off (at the upper end) at $Q \sim Q_2 \sqrt{U_0/U}$.  
Hence we get $\ln [(Q_2/Q_1) \sqrt{U_0/U}]$. 
With $U \sim U_0$ for $Q \in (Q_1, Q_2)$, this is $\sim \ln(Q_2/Q_1)\sim \ln(1/\Lambda)$. 
The remaining integral over $U$ can be estimated as $\sim U_0^4$. 
Thus we obtain 
\beq
 P_{\rm Inflation} \sim v_0 \Lambda^6 N_{\rm CMB}^{-2} \ln (1/\Lambda^2). 
\eeq
This is consistent with our estimate in Ref.~\cite{Masoumi:2016eag} if we take the volume factor 
$f_{\rm vol} \sim v_0 \Lambda^2$ into account. 
This probability is quite small if we require $N_{\rm CMB} \gtrsim 50$ 
and assume $\Lambda \ll 1$. 
However, once we impose an anthropic constraint on the total e-folding number, 
the conditional probability of having $N_e \ge N_{\rm CMB}$ will be as large as unity. 
We will discuss the anthropic conditions in Sec.~\ref{sec:observation}.

The probability distribution is plotted in Fig.~\ref{fig:P} 
as a function of $Q$, where we assume inflection point inflation with 
$N_{\rm max}= 120$, $N_{\rm CMB} = 50$, $U_0 = 10^{-19}$, $\bar{U} = 0$, and 
$\Lambda = 0.1$ (blue curve) or $0.02$ (orange curve). 
For these parameter values the spectral index of density perturbations 
is $n_s \simeq n_s^{(\rm obs)} \simeq 0.97$.  We calculated the curves 
in Fig.~\ref{fig:P} directly from the integral (\ref{PNQ}), which can be evaluated 
analytically (with a somewhat unwieldy result).  They agree very well with our approximate 
power law expressions (\ref{PNQ<}), (\ref{PNQ<>}), (\ref{PNQ>}) in the 3 different
regimes divided by $Q_1 < Q< Q_2$.

\begin{figure}[t] 
   \centering
   \includegraphics[width=5in]{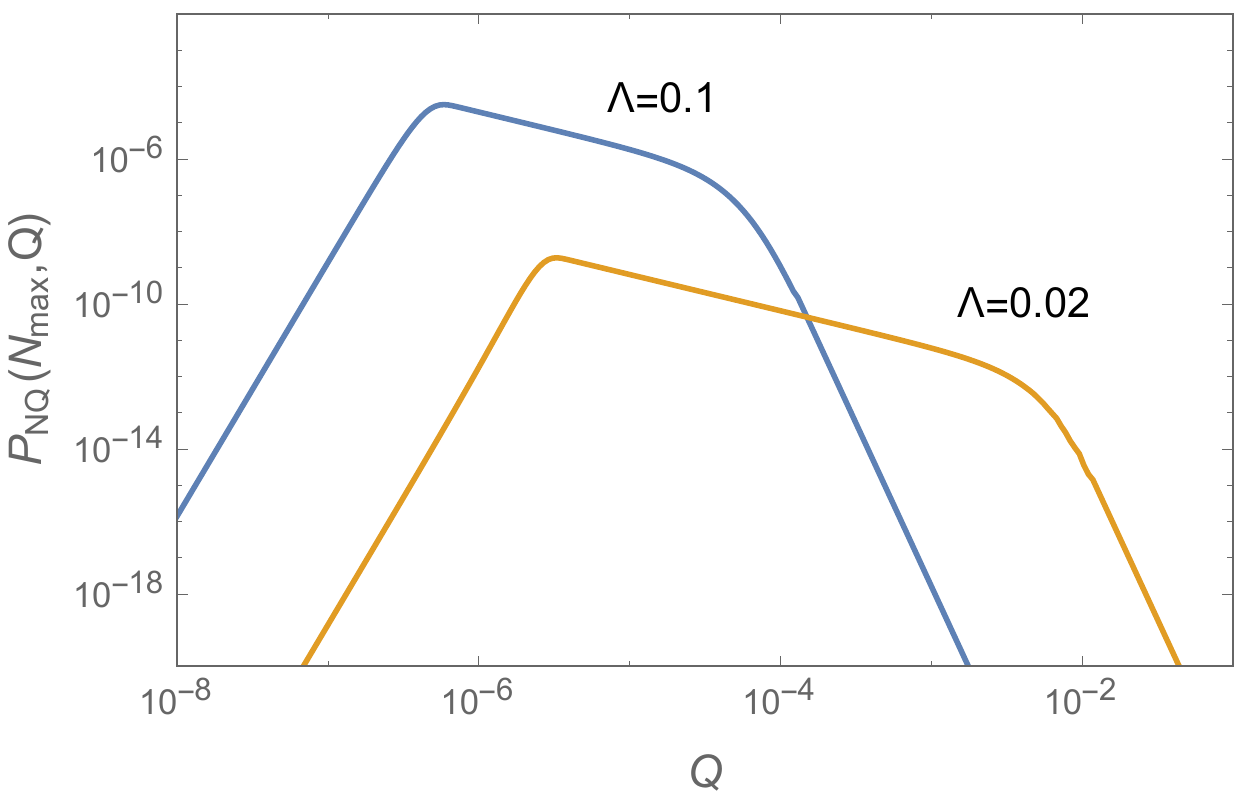} 
   \caption{
 $P_{NQ} (N_{\rm max},Q)$ as a function of $Q$. 
 We assume $N_{\rm max} = 120$, $N_{\rm CMB} = 50$, $U_0 = 10^{-19}$, $\bar{U} = 0$, 
and $\Lambda = 0.1$ (blue curve) or $0.02$ (orange curve). 
   }
   \label{fig:P}
\end{figure}

In summary, the prior distribution
(\ref{PNQ}) together with the condition (\ref{condition}) is the main result of this section. 
It can be written as 
$P_{NQ}(N_{\rm max},Q) \propto N_{\rm max}^{-3} P(Q)$
with
\beq
&P(Q)\propto Q^{7} &\qquad {\rm for} \ Q < Q_1
\\
&P(Q)\propto Q^{-1}  &\qquad {\rm for} \ Q_1 < Q < Q_2
\\
&P(Q)\propto Q^{-9}  &\qquad {\rm for} \ Q_2 < Q, 
\label{PQ}
\eeq
where $Q_1$ and $Q_2$ are defined in Eqs.~(\ref{condition}) and 
(\ref{R}), respectively.  
Substituting this into \eq{PnP}, we can calculate the probability distribution for observables 
in the multiverse, which will be discussed in Sec.~\ref{sec:observation}. 
In the next section, we will show that \eq{PNQ} is still correct with some minor modifications in a higher-dimensional landscape.

\section{A higher-dimensional Landscape}
\label{sec:multifield}

We now consider the probability distribution of $N_{\rm max}$ and $Q$ in a $D$-dimensional 
field space.  We will see that the calculation reduces to the one-dimensional case with some minor modifications. 

In a multi-field landscape, the Taylor expansion of the potential can be written as
\beq
U({\bm\phi})=U+\eta_i\phi_i+\frac{1}{2}\zeta_{ij}\phi_i\phi_j+\frac{1}{3!}\rho_{ijk}\phi_i\phi_j\phi_k + ...,
\eeq
where $i,j,k = 1,2, ... , D$.  The expansion coefficients are $\eta_i=\partial U/\partial \phi_i$, $\zeta_{ij} =
\partial U/\partial \phi_i\partial\phi_j$ and $\rho_{ijk}=\partial U/\partial \phi_i\partial\phi_j\partial\phi_k$, 
with all derivatives taken at ${\phi}_i=0$.  Their typical values in the landscape are 
$\eta_i\sim U_0/\Lambda$, $\zeta_{ij}\sim U_0/\Lambda^2$, $\rho_{ijk}\sim U_0/\Lambda^3$.

Multi-field analogues of saddle points and inflection points can be defined as follows.  A saddle 
point is a point where ${\del}_i U=0$ and the Hessian matrix $\zeta_{ij}$ has one negative 
eigenvalue, with all other eigenvalues positive.  An inflection point is a point where one of the 
Hessian eigenvalues is zero, with all other eigenvalues positive, and the gradient of $U({\bm\phi})$ 
vanishes in all directions orthogonal to that of the zero eigenvalue.  Inflation is also possible at 
points with several negative or zero eigenvalues, but this occurs very rarely in a small-field 
landscape \cite{Masoumi:2017xbe}.  In the rest of the paper we shall disregard this possibility.

\subsection{Inflection point inflation}

We first consider a low-energy landscape with a fixed value of the average potential ${\bar U}$.  
Integration over ${\bar U}$ in Eq.~(\ref{prob}) will be performed later.
Ensemble averages over inflection (or saddle) points in the landscape
can be calculated by integrating over ${\bm \phi}$ 
with appropriate delta functions $\prod_i \delta (f_i ({\bm \phi}))$. 
Without loss of generality, we can diagonalize the Hessian, $\zeta_{ij}=\lambda_i \delta_{ij}$, 
and choose the $\phi_1$ axis in the direction of zero (or negative) eigenvalue.  The slow roll will 
then occur essentially along the $\phi_1$ axis.  To simplify the equations, we shall use the notation
$\lambda \equiv \lambda_1$, $\eta \equiv \eta_1$, and $\rho\equiv\rho_{111}$. 

Since for inflection points we require 
$\lambda=0$, $\lambda_a \ge 0$, and $\eta_a = 0$ for $a =2,3, \dots,D$, 
we set $f_1({\bm \phi}) = \lambda$ and $f_a ({\bm \phi}) = \eta_a$. 
The Jacobian associated with the delta functions is then given by $\abs{ \rho 
\prod_a \lambda_a}$.  Hence the probability that inflection-point inflation with certain values 
of $N_{\rm max}$ and $Q$ will occur starting from a randomly chosen point in a landscape of 
average energy ${\bar U}$ is given by 
\beq
 &&\hspace{-0.6cm} P_{NQU} (N_{\rm max}, Q, \bar{U}) 
 = \frac{D}{V} 
 \int d^D \phi 
 dU  \prod_i d\eta_i \prod_i  d \lambda_i \prod_{ijk} d\rho_{ijk} 
J(\lambda_i)
{\cal P} (U, \eta_i, \lambda_i, \rho_{ijk}) f_{\rm vol} (U, \rho) ~ \delta(\lambda)|\rho|\times
\nonumber\\
&&
 \hspace{-0.6cm}\times \left( \prod_{a=2}^D \delta(\eta_a) |\lambda_a| \right) 
 \lmk \prod_{a=2}^D \theta \lmk \lambda_a \rmk 
\rmk
\theta \lmk \eta \rho \rmk 
\theta (U) 
\delta\left(N_{\rm max} -\frac{\pi\sqrt{2} U}{\sqrt{\eta\rho}}\right) 
\delta\left(Q-\frac{N_{\rm CMB}^2 \rho}{4 \sqrt{3\pi} U^{1/2}} f(x,y) \right)
\label{PQNU2}
\eeq
where we have included the volume factor $f_{\rm vol}\sim v_0 U/(\Lambda \rho)$ . 
The combinatorial factor $D$ comes from selecting $\lambda_1$ to be the smallest eigenvalue. 
The integral $\int d^D \phi$ gives a volume in the field space $V$ 
because of the homogeneity of probability distribution ${\cal P}$. 
The Jacobian $J(\lambda_i)$ comes from the variable transformation from $\zeta_{ij}$ to $\lambda_i$: 
\beq
 J(\lambda_i) = C \prod_{i \ne j} \abs{\lambda_i - \lambda_j}, 
\label{J}
\eeq
where $C$ is a constant. Finally the terms $\theta \lmk \eta \rho \rmk 
\theta (U)$ are included to ensure that slow roll is possible and does not lead to a shallow minimum.
The distribution
${\cal P} (U, \eta_i, \lambda_i, \rho_{ijk})$ has the form
\beq
&& {\cal P} (U, \eta_i, \lambda_i, \rho_{ijk})  = {\cal P}_1 (U, \lambda_i) {\cal P}_2 (\eta_i, \rho_{ijk}) 
\\
 &&{\cal P}_1 = A_1 \exp [ -{\cal Q}_1(U,\lambda_i)] 
 \\
 &&{\cal P}_2 = A_2 \exp [ -{\cal Q}_2(\eta_i,\rho_{ijk})], 
\eeq
where $A_1$ and $A_2$ are determined by the normalization conditions: 
\beq
 &&\int d U \prod_i d \lambda_i {\cal P}_1 (U, \lambda_i)  = 1
 \\
 &&\int \prod_i d \eta_i \prod_{ijk} d \rho_{ijk} {\cal P}_2 (\eta_i, \rho_{ijk}) = 1.
 \label{P_2 normalization}
\eeq
The exponents ${\cal Q}_1(U,\lambda_i)$ 
and ${\cal Q}_2(\eta_i,\rho_{ijk})$ are given by~\cite{Fyodorov,Bray:2007tf,Masoumi:2016eag}
\beq
&&{\cal Q}_1(U,\lambda_{i}) = 
c_1 \frac{1}{U_0^2} (U - \bar{U})^2 - c_2 \frac{\Lambda^2}{D U_0^2} (U - \bar{U}) \sum_i \lambda_i  - 
c_3 \frac{\Lambda^4}{D U_0^2} \lmk \sum_i \lambda_i \rmk^2   +c_4 \frac{\Lambda^4}{U_0^2} \lambda_i \lambda_i 
\label{Q1} 
\nonumber
\\
&&{\cal Q}_2(\eta_i,\rho_{ijk}) = 
c_5 \frac{\Lambda^2}{U_0^2} \eta_i \eta_i +c_6 \frac{\Lambda^4}{DU_0^2} \eta_i \rho_{ijj}  - 
c_7 \frac{\Lambda^6}{DU_0^2} \rho_{iik}\rho_{jjk} +c_8 \frac{\Lambda^6}{U_0^2} \rho_{ijk}\rho_{ijk},
\label{Q2}
\eeq
with summation over repeated indices. 
The coefficients $c_i$ can be expressed in terms of the moments of the correlation function 
and are typically $\mathcal{O}(1)$. 
\eq{PQNU2} can be rewritten as 
\beq
 &&P_{NQU} (N_{\rm max}, Q, \bar{U}) 
 = 
 D \int 
 dU  d \eta d \lambda d \rho ~
{\cal P}_1' (U, \lambda; \bar{U})~ 
 {\cal P}_2' (\eta, \rho) ~
|\rho| 
\delta (\lambda) ~
f_{\rm vol} (U, \rho) ~
\theta (U) 
\theta \lmk \eta \rho \rmk 
\nn
&&~~~~~~~~~~~~~~~~~~~~~~~~\times ~
\delta\left(N_{\rm max} -\frac{\pi\sqrt{2} U}{\sqrt{\eta\rho}}\right) 
\delta\left(Q-\frac{N_{\rm CMB}^2 \rho}{4 \sqrt{3\pi} U^{1/2}} f(x,y) \right), 
\label{factorizedP}
\eeq
where
\beq
&&{\cal P}_1' (U, \lambda; \bar{U})  = 
\int \prod_{a = 2}^D d \lambda_a 
{\cal P}_1 (U, \lambda_i)  J(\lambda_i) 
 \prod_{a=2}^D |\lambda_a| \theta \lmk \lambda_a \rmk ,
\label{P1'} 
\\
&&{\cal P}_2'(\eta, \rho) = 
\int \prod_{a =2}^D d \eta_a \prod_{(ijk) \ne (111)} d \rho_{ijk}
{\cal P}_2 (\eta_i, \rho_{ijk}) 
\left( \prod_{a=2}^D \delta(\eta_a) \right). 
\label{P_2p}
\eeq

The delta function $\delta(\lambda)$ sets $\lambda=0$ in the distribution (\ref{P1'}), 
while the other eigenvalues are integrated over. The resulting expression can be found using the 
saddle point method of Ref.~\cite{Bray:2007tf}; it is given by 
\beq
 {\cal P}_1'(U, 0 ;\bar{U} ) \sim 
 \frac{U_0^{D-3}}{\Lambda^{2 D-4}} 
 \exp \lkk - \lmk d_1' \frac{U - \bar{U}}{U_0} 
 + d_2' \sqrt{D} 
\rmk^2 - d_3' \lmk \frac{U - \bar{U}}{U_0} \rmk^2 \rkk, 
 \label{P_1p}
\eeq
where $d_i'$ 
are ${\cal O}(1)$ constants.  A detailed derivation of (\ref{P_1p}) will be discussed elsewhere~\cite{Yamada:2017uzq}. 

The exponent of ${\cal P'}_2$ is quadratic in all variables, 
so we can integrate $\eta_a$ ($a \ge 2$) and $\rho_{ijk} ((i,j,k) \ne (111))$. 
We find that 
the coefficients of the remaining terms that are proportional to $\eta^2$ or $\rho^2$ 
do not change after the integration in the large $D$ limit. This is explained in detail in Appendix~\ref{sec:quadratic}.  
As a result, we have 
\beq 
 {\cal P}_2' (\eta, \rho) 
 \sim 
 \frac{\Lambda^{D+3}}{U_0^{D+1}} 
 \exp \lkk - c_5 \frac{\Lambda^2}{U_0^2} \eta^2 - 3c_8 \frac{\Lambda^6}{U_0^2} \rho^2 \rkk . 
\eeq

We are now ready to perform the integration over ${\bar U}$ in Eq.~(\ref{prob}). In order to do that
we first note that the dependence on  ${\bar U}$ in $P_{NQU} (N_{\rm max}, Q, \bar{U})$ is captured 
exclusively by ${\cal P}_1'(U, 0 ;\bar{U} )$, so we just need to compute, 
\beq
 &&{\cal P}_1'' (U) =  \int d \bar{U} P_{\rm cc} (0; \bar{U}) 
 {\cal P}_1'(U, 0; \bar{U} ) \sim 
 \frac{\Lambda^4}{U_0^2} \frac{U_0^{D-1}}{\Lambda^{2 D}} 
 \exp \lkk 
 - c_1' \frac{ U^2}{U_0^2} 
 \rkk P' ,
 \label{P1''}
\\
 &&P' \sim  \exp \lkk - d_3''^2 D \rkk, 
\eeq
where $c_1'$, $d_3'' = \mathcal{O}(1)$. The exponential suppression factor $P'$ is 
related to the fact that stationary points of index $(D-1)$ (that is, 
having all but one Hessian eigenvalues positive) are rather rare in the landscape.

After integration over $\eta$ and $\rho$, we obtain the following expression for the distribution $ P_{NQ}$, 
which was defined in Eq.~(\ref{prob}):
\beq
 P_{NQ} (N_{\rm max}, Q) 
 &=&  \int d \bar{U}  P_{\rm cc} (0, \bar{U}) P_{NQU} (N_{\rm max}, Q, \bar{U}) 
 \\
 &=& 
 D \frac{4 \pi^{5/2} v_0 }{Q N_{\rm max}^3} 
\int dU  
 U^{3} \theta (U) 
 {\cal P}_1'' (U) {\cal P}_2' (\eta_*, \rho_*), 
 \label{PI}
\eeq
where $\eta_*$ and $\rho_*$ are the values (\ref{*0}) selected by the delta functions. 

As before, $\eta_*$ is much smaller than the typical value $\eta_0 \sim U_0/\Lambda$, so we can set 
$\eta_*\approx 0$ in the exponent of ${\cal P}_2'$.  We thus obtain
\beq
  P_{NQ} (N_{\rm max}, Q) 
  \sim 
  \frac{P'}{\Lambda^D} 
 \frac{D v_0 \Lambda^7}{U_0^4} 
\frac{1}{Q N_{\rm max}^3} 
 \int_0^\infty dU  
 U^{3} 
\exp \lkk 
- c_1' \frac{U^2}{U_0^2} 
 - 3c_8 \frac{\Lambda^6}{U_0^2} \rho_*^2 
 \rkk, 
\label{result}
\eeq
We note that the factor $P' / \Lambda^{D}$ is roughly the density of index-$(D-1)$ inflection points in the landscape.

Eq.~(\ref{result}) is very similar to \eq{PNQ} for the 1D case with ${\bar U}=0$.  The difference is in the constant 
pre-factor and in ${\cal O}(1)$ coefficients in the exponent.  Therefore, our results for the distribution of Q in 
Eqs.~(\ref{PNQ<})-(\ref{PNQ>}) should apply with these corrections.%

In this Section we assumed, following the analysis in Ref.~\cite{Masoumi:2017xbe}, that inflation in a multi-dimensional landscape is essentially single-field.  This requires that all eigenvalues $\lambda_j$ of the Hessian matrix $\zeta_{ij} = \partial^2 U/\partial \phi_i \partial\phi_j$ at the inflection point, apart from one zero eigenvalue, are large compared to the inflaton potential $U$.  The typical value $\zeta_{ij}\sim U/\Lambda^2$ is much larger than $U$ for small values of $\Lambda$.  In a higher-dimensional landscape, $D\gg 1$, some of the eigenvalues $\lambda_j$ may by chance be small.  But for a sufficiently small $\Lambda$ we expect $1D$ inflation to be statistically prevalent.
A careful analysis in Ref.~\cite{Yamada:2017uzq} shows that this condition is satisfied if
\beq
\Lambda \ll D^{-1/4}, 
\label{LambdaN}
\eeq
For $D \sim 100$, the condition is $\Lambda\ll 0.3$.

We note that inflation in random Gaussian models has been recently studied by Bjorkmo and Marsh \cite{Bjorkmo:2017nzd}, who concluded that multi-field effects are generically important in such models.  The reason for this discrepancy is that the landscape properties assumed in Ref.~\cite{Bjorkmo:2017nzd} are significantly different from ours.  The most significant difference is that they assume that the correlation function has a Gaussian form, $F(x)\propto \exp(-x^2/2)$.   This is a rather special case, in which the distribution of Hessian eigenvalues at a stationary point has a sharp peak around zero, so it is not surprising that some fields are light enough to be excited during inflation~\cite{Yamada:2017uzq}.  On the other hand, we assumed a generic correlation function, which does not have this property.
Moreover, in most of their numerical examples Bjorkmo and Marsh used the parameters $\Lambda\ = 0.4$ and $N\leq 100$, which do not satisfy the condition (\ref{LambdaN}).

\subsection{Saddle point inflation}

For saddle point inflation, the analogue of the distribution (\ref{PQNU2}) is
\beq
&&\hspace{-0.6cm}
  P_{NQU} (N_{\rm max}, Q, {\bar U}) 
 = \frac{D}{V } 
 \int \dots \int d^D \phi 
 dU  \prod_i d\eta_i \prod_i  d \lambda_i \prod_{ijk} d\rho_{ijk} 
J(\lambda_i)
{\cal P} (U, \eta_i, \lambda_i, \rho_{ijk}) f_{\rm vol} (U, \rho) 
\nonumber\\
&&\hspace{-0.4cm}
\times \delta(\eta)|\lambda| \left( \prod_{a=2}^D \delta(\eta_a) |\lambda_a| \right) 
 \lmk \prod_{a=2}^D \theta \lmk \lambda_a \rmk 
\rmk
\theta \lmk - \lambda \rmk 
\theta (U) 
\delta\left(N_{\rm max} -\frac{2 \pi U}{\abs{\lambda}}\right) 
\delta\left(Q-\frac{N_{\rm CMB}^2 \rho }{4 \sqrt{3\pi} U^{1/2}} 
f(x,y) \right), 
\nonumber
\eeq
where $f_{\rm vol} (U, \rho) \sim U / \rho$. 

The delta functions fix $\lambda$ and $\rho$ to the values
\beq
 \lambda_* (U) = \frac{2\pi U}{N_{\rm max}},
 ~~~~
 \rho_* (U) = \frac{4 \sqrt{3 \pi} Q U^{1/2}}{N_{\rm CMB}^2 f(x,y)}. 
\eeq
The slow roll condition requires $\lambda_* < U \lesssim U_0$, which is much smaller than the typical value 
of $\lambda$.  Hence we can set $\lambda_* \approx 0$ in the exponent of ${\cal P}$. 

The remaining integrals can be evaluated following the same steps as in the preceding subsection, with only 
minor changes.  For example, ${\cal P}_1'(U, 0;\bar{U} )$ is now replaced with ${\cal P}_1'(U, \lambda_*;\bar{U} )$.  
This change, however, has little effect on the distribution.  It was shown in Ref.~\cite{Yamada:2017uzq} that with 
$\lambda_1\approx 0$, the second smallest eigenvalue of the Hessian is of the order 
$\lambda_2 \sim U_0 / (\sqrt{D} \Lambda^2)$. This is much greater than $\lambda_*$ if 
$\Lambda \ll D^{-1/4} \approx 0.3$ (for $D\sim 100$).  Then the Jacobian $J(\lambda_i)$ in Eq.~(\ref{J}) 
changes very little when we replace $\lambda_1=0$ by $\lambda_*$, so we can estimate 
\beq
 {\cal P}_1'(U, \lambda_*;\bar{U} ) \sim 
 {\cal P}_1'(U, 0;\bar{U} ). 
\eeq

After integration over $\bar{U}$, we finally obtain 
\beq
  P_{NQ} (N_{\rm max}, Q) 
 &\sim& 
 D \frac{1}{\sqrt{2 \pi}} \frac{4 \pi^{5/2}}{Q N_{\rm max}^3} 
 \int_0^\infty d U U^{3}  
  {\cal P}_1'' (U) 
  {\cal P}_2' (0, \rho_*) 
  \\
  &\sim&
  \frac{P'}{\Lambda^D} 
 \frac{D \Lambda^7}{U_0^4} 
\frac{1}{Q N_{\rm max}^3} 
 \int_0^\infty dU  
 U^{3} 
\exp \lkk 
- c_1' \frac{U^2}{U_0^2} 
 - 3c_8 \frac{\Lambda^6}{U_0^2} \rho_*^2 
 \rkk. 
\label{result2}
\eeq
This is the same as \eq{result} but 
without a factor of $v_0$ and 
with a different $f(x,y)$ in $\rho_*(U)$ (see Appendix A). 
Note that $f(x,y) \simeq 1$ for $N_{\rm max} \gtrsim N_{\rm CMB}$ 
and the difference is small in this case, because the observable scale leaves the
horizon when the potential is dominated by the cubic term. Thus the probability of saddle point inflation is suppressed by a factor of $1/v_0$ ($\sim 0.06$) compared with 
that of inflection point inflation with the same $N_{\rm max}$ and $Q$.

\section{Observational predictions}
\label{sec:observation}

Our main result is that the ``prior" probability distribution in Eq.~(\ref{PnP}) has the form
\beq
P_{NQ}(N_{\rm max},Q) \propto N_{\rm max}^{-3} P(Q)
\label{prior}
\eeq
with
\beq
P(Q)\propto Q^{-1}.
\label{PQ}
\eeq
It applies in the range $Q_1<Q<Q_2$, where $Q_1$ and $Q_2$ are defined in Eqs.~(\ref{condition}) and 
(\ref{R}), respectively.  Outside of this range, $P(Q)$ rapidly declines towards zero.  

To derive observational predictions, we also need to know the anthropic factor 
$n^{\rm (obs)}(N_{\rm max},Q,\rho_v)$ in Eq.~(\ref{PnP}).  We will not attempt a detailed analysis 
here and will only give a rough outline of the observational implications of Eqs.~(\ref{prior}),(\ref{PQ}).

"Pocket universes" resulting from bubble nucleation have negative spatial curvature, which depends on the number of e-folds of inflation in the bubble, $N_e\lesssim N_{\rm max}$.  The $N_{\rm max}$ 
dependence in Eq.~(\ref{prior}) disfavors large values of $N_{\rm max}$ and thus favors large magnitude of the curvature.  On the other hand, if the curvature is so large that it suppresses structure formation, the density of observers is also suppressed. 
This gives an anthropic upper bound on the magnitude of curvature and a lower bound on $N_e$  \cite{Vilenkin:1996ar,Freivogel:2005vv}.  It was noted in Ref.~\cite{Freivogel:2005vv} that these bounds are rather close to the observational bounds.  For example, in models where inflation is at a GUT scale and thermalization is instantaneous, the anthropic bound is $N_e \gtrsim 60$, while the observational bound is $N_e > 62$.  This is consistent with the distribution (\ref{prior}), which suggests that $N_e$ should be close to the smallest anthropically allowed value.\footnote{Note that $N_e$ is typically $\sim N_{\rm max}$ for inflection point inflation \cite{Masoumi:2017gmh}.}

Anthropic bounds on the amplitude of density perturbations $Q$ have been discussed in Refs.~\cite{Tegmark:1997in,Tegmark:2005dy}, with the conclusion that both lower and upper bounds are within an order of magnitude of the observed value $Q^{\rm (obs)} \sim 10^{-4}$. 
If $Q^{\rm (obs)}$ were smaller than $Q_1$ or larger than $Q_2$, it would be in the range where 
$P(Q)$ changes very rapidly.  The expected value of $Q$ would then be pushed into the anthropically unfavorable territory, so finding $Q\sim Q_{\rm obs}$, which is comfortably within the anthropically favored zone, would be rather unlikely.
We thus conclude that the observed 
value of $Q\sim 10^{-4}$ should be within the range of $(Q_1, Q_2)$, 
which requires 
\beq
 \frac{U_0}{\Lambda^2} \lesssim 10^{-13} \lesssim 
 \frac{U_0}{\Lambda^6}.
 \label{conditionQ}
\eeq
This gives a restriction on observationally acceptable models of random Gaussian landscape.

Assuming $Q_1 < Q^{\rm (obs)} < Q_2$, 
we next estimate 
the anthropic factor 
$n^{\rm (obs)}(N_{\rm max},Q,\rho_v)$ in more detail. 

The observer density $n^{\rm (obs)}$ is expected to be roughly proportional to the fraction of matter 
$f_G$ clustered in large galaxies (with mass $M\gtrsim 10^{11} M_\odot$).\footnote{Smaller galaxies lose much 
of their baryons due to the wind from supernova explosions.}  The idea is that there is a certain number of stars per 
unit mass in a galaxy and certain number of observers per star.  The mass 
fraction $f_G$ can be found in the Press-Schechter approximation.  Restricting attention to positive values 
of $\rho_v$ and assuming that $N_{\rm max}$ is large enough to yield a nearly flat universe, it is given 
by \cite{Martel:1997vi, Garriga:2002tq}
\beq
f_G \propto {\rm erfc} \left[0.8 \left(\frac{\rho_v}{\rho_m \sigma_G^3}\right)^{1/3}\right],
\label{fG}
\eeq
where $\rho_m$ is the density of nonrelativistic matter and $\sigma_G$ is the linearized density contrast 
on the galactic scale.  $\sigma_G$ is linearly related to the primordial fluctuation amplitude $Q$, 
$\sigma_G\propto Q$.  The product $\rho_m \sigma_G^3$ is time-independent during the matter era 
and can be evaluated at any time.  

From Eqs.~(\ref{PQ}) and (\ref{fG}), the combined probability distribution for $\rho_v$ and $Q$ in the 
multiverse can be written as
\beq
dP(\rho_v,Q) \propto \frac{dQ}{Q} {\rm erfc} (0.8\xi^{1/3})d\rho_v,
\label{dP}
\eeq
where 
\beq
\xi \equiv \frac{\rho_v}{\rho_m \sigma_G^3} \propto \frac{\rho_v}{Q^3}.  
\eeq
With a change of variables $\{Q,\rho_v\}\to\{Q,\xi\}$, 
this distribution factorizes \cite{Garriga:1999hu}:
\beq
dP(\xi,Q) \propto Q^2 {dQ} \times {\rm erfc} (0.8\xi^{1/3})d\xi
\label{dP2}
\eeq

The distribution for $\xi$ is peaked at $\xi\sim 1$ (on a logarithmic scale of $\xi$).  With $Q\sim 10^{-4}$, 
the corresponding value of $\rho_v$ is comparable to the observed value $\rho_v^{(0)}$.  But more generally, 
this distribution predicts the value of $\rho_v/Q^3$. 

The novel aspect of Eq.~(\ref{dP2}) is the distribution for $Q$.  This distribution applies in the range from 
$Q_1$ to $Q_2$; hence it is peaked at $Q_2$.  

In the above analysis we made a number of simplifications, which we shall now spell out.

{\it (i)} $f_G$ in Eq.~(\ref{fG}) is the {\it asymptotic} mass fraction, while in the scale factor measure we need 
to use the density of observers at a finite time $t_{\rm obs}$.  This distinction, however, has little effect on 
the probability distribution.  

{\it (ii)} In the scale-factor measure, the distribution (\ref{dP}) has an additional factor $\sim \exp(-3H_v t_{\rm obs})$, 
which arises due to the change in the expansion rate after vacuum energy domination \cite{DeSimone:2008bq,Bousso:2008hz}.  Here,  
$H_v=(\rho_v/3)^{1/2}$ is the expansion rate during the vacuum dominated epoch.  This factor suppresses 
large values of $\rho_v$, but its effect is not very significant for observationally interesting values of 
$\rho_v$ and $t_{\rm obs}$.  

{\it (iii)} The density of observers can be influenced by a number of other factors that we have not included 
here.  For example, regions with large values of $Q$ may be disfavored due to life extinctions caused by 
close encounters with stars or molecular clouds \cite{Tegmark:1997in}.  If the dangerous value $Q_*$ above 
which this effect is significant is $Q_*<Q_2$, we can expect to observe $Q\sim Q_*$, which means that 
the rate of extinctions is close to the dangerous level.  This prediction is consistent with the fact that 
great extinctions on Earth occurred once in $\sim 10^8$ years, which is about the time that it took 
intelligent life to evolve.\footnote{For $Q_* < Q_1$ this argument would also suggest an observed value $Q\sim Q_*$.  However, the dependence $P(Q)\propto Q^9$ is very steep and would probably push the predicted value too far into the dangerous range.}
Life extinctions could also be caused by gamma-ray bursts.  This could suppress
 the probability of very small and negative values of $\rho_v$ \cite{Piran:2014wfa}.

\section{Conclusions and discussion}\label{sec:conclusion}

The idea that the Landscape of String Theory can be composed of 
more than one sectors of the moduli space with different energy scales motivates 
the study of the cosmological implications of a random landscape with this
structure. In this paper we have assumed that there are two distinct
sectors of the landscape with a hierarchy of energies, a low energy and a high-energy scale. 
This type of model captures some of the key features of string theory compactifications, 
however, it is still a toy model and the specific form of the potential  found in the literature 
may be different from the one we present here. Nevertheless we believe the conclusions 
of this paper will be also applicable for more realistic models of compactification.

With this structure of the landscape, we have shown that the initial conditions
for our primordial universe are likely to be determined by quantum tunneling between the vacua 
in the field space of the high energy sector. Moreover, the field values 
of the low energy moduli sector do not change much due to this quantum
tunneling process. Since we do not expect strong correlations between 
the potentials of these fields,  the initial values of the fields in the
low energy sector would have a flat distribution. The subsequent cosmological
evolution inside the bubble is mostly determined by the dynamics of the low
energy landscape sector. Inflationary periods in this sector would be
dominated by trajectories that fall within the attractor region
of fine tuned inflection points and saddle points of the landscape.

This picture allowed us to compute 
the probability distribution for the (maximal) number of e-folds $N_{\rm max}$,
the amplitude of scalar fluctuations $Q$, 
and the vacuum energy density $\rho_v$ in the multiverse for this model. The resulting distribution can be represented as
\beq
dP \propto P_{\rm prior} n_{\rm obs} dN_{\rm max} dQ d\rho_v,
\eeq
where the prior distribution has the form
\beq
P_{\rm prior} (N_{\rm max},Q) \propto N_{\rm max}^{-3} P(Q)
\label{prioragain}
\eeq
and is (to a good approximation) independent of $\rho_v$.
We found that the distribution for $Q$ in Eq.~(\ref{prioragain}) is $P(Q)\propto Q^{-1}$ in the range $Q_1 <Q<Q_2$, with $Q_1$ and $Q_2$ respectively given by Eqs.~(\ref{condition}) and 
(\ref{R}), and drops rapidly towards zero outside of this range.  
Requiring that the observed value of $Q$ falls between $Q_1$ and $Q_2$, we obtained a constraint on the model parameters, Eq.~(\ref{conditionQ}).  
We also found that the probability of saddle point inflation is smaller than that of inflection point inflation, roughly by an order of magnitude.

The prior distribution specified above is the main result of this paper.  We also briefly discussed the anthropic factor $n^{(\rm obs)} (N_{\rm max},Q,\rho_v)$.  The number of e-folds $N_{\rm max}$ should be large enough, so that curvature does not dominate prior to galaxy formation.  Since $P_{\rm prior}$ is a decreasing function of $N_{\rm max}$, we expect that the observed value of $N_{\rm max}$ should be comparable to this anthropic bound.  Disregarding the $N_{\rm max}$-dependence, the full distribution for $\rho_v$ and $Q$ (including the anthropic factor) can be represented in a factorized form
\beq
dP(\xi,Q) \propto Q^2 {dQ} ~ F(\xi) d\xi,
\label{factor}
\eeq
where $\xi \propto \rho_v/Q^3$.  The new variable $\xi$ and the function $F$ are specified in Sec.~\ref{sec:observation}.  The value of $\xi$ corresponding to the observed value of the combination 
$\rho_v/Q^3$ is close to the peak of the function $F(\xi)$.  This manifests the celebrated anthropic solution of the cosmological constant problem.  The novel aspect of Eq.~(\ref{factor}) is the distribution for $Q$, which applies in the range $Q_1<Q<Q_2$.

While we used a random Gaussian model to describe the low-energy landscape $U_{\rm L}({\bf \phi},{\bf \psi})$, our conclusions are rather insensitive to the assumptions about the high-energy landscape $U_{\rm H}({\bf \psi})$. 
We only used three assumptions regarding the high-energy landscape: (i) that the tunneling rate in the potential $U_{\rm H}({\bf\psi})$ is much higher than that in $U_{\rm L}$, (ii) that the tunneling points are uncorrelated with the low-energy potential, and (iii) that the distribution of vacuum energy densities in $U_{\rm H}$ is much wider than $U_{\rm L}$.  These assumptions are likely to be satisfied in a very wide class of models.

Our conclusions may also be applicable to one-scale models with $U_{\rm H} = U_{\rm L}$ under an additional condition. In our calculations, we used the facts that the initial condition for slow-roll inflation has a flat distribution in the low-energy landscape and that the volume of the attractor region $f_{\rm vol}$ is independent of the dimension of landscape. These are true even for one-scale models with $U_{\rm H} = U_{\rm L}$ if the dependence of the tunneling rate on the tunneling point in the attractor region is negligible. In this case, the tunneling probability is peaked along the slow-roll direction~\cite{Masoumi:2017xbe} and the length of the slow-roll track is $(v_0 U/\rho)$, which is parametrically equal to $f_{\rm vol}$. The condition can then be written as $1 \gg (v_0 U/\rho) \cdot dS/d\phi$. 
The factor of $dS/d\phi$ can be roughly estimated as $(\Lambda^4 / U) \bar{S} / \Lambda$ from Eq.~(\ref{Sij}), where $\bar{S} \gtrsim {\cal O}(10)$. 
The parameter $\rho$ can be rewritten in terms of $Q$ and $U^{1/2}$ from Eq.~(\ref{*0}) 
and $U$ is $\sim U_0$ in the anthropically allowed range we are interested in. 
As a result, the condition can be rewritten as $U_0/\Lambda^6 \gg v_0^2 f^2 N_{\rm CMB}^4 \bar{S}^2/(48 \pi Q^2) \sim 10^{17}$ for $Q \sim 10^{-4}$. Hence our conclusions are applicable also to one-scale models with $U_0/\Lambda^6 \gg 10^{17}$.

The methods we used here can be applied to other models, in particular to the random $\alpha$-attractor model recently introduced in Ref.~\cite{Linde:2016uec}.
In this model some directions have a flat potential due to a singularity of their kinetic terms,
so that the effective mass in those directions is
much smaller than that in the perpendicular directions. 
Some of our results and methods may also be applicable to axionic landscapes 
which have been studied recently in Ref.~\cite{Bachlechner:2017hsj}.

\section*{Acknowledgments}
J.J. B.-P. would like to thank useful conversations with Kepa Sousa and Mikel Alvarez. This work 
is supported in part by the Basque Foundation for Science (IKERBASQUE), the 
Spanish Ministry MINECO  grant (FPA2015-64041-C2-1P), the  Basque Government grant (IT-979-16)
and the National Science Foundation under grant 1518742. J.J. B.-P. would also like to thank the Tufts 
Institute of Cosmology for its kind hospitality during the time
that this work was completed. 

\nocite{}

\appendix

\vspace{1cm}

\section{Observables of Inflation}
\label{sec:observables}

In this Appendix we review the derivation of 
spectral index $n_s$ 
and 
amplitude of scalar fluctuations $Q$ 
in inflection point inflation and saddle point inflation. 

\subsection{Inflection point inflation}

The potential for inflection point inflation is
\beq
U(\phi)=U + \eta\phi + \frac{1}{3!}\rho \phi^3 +..., 
\eeq
where $\eta \rho > 0$. 
The slow roll ends when $|U''|/U = 1$, or
\beq
\phi_{\rm end} = -U/ \rho.
\eeq
The number of e-folds from the observable scale $\phi$ to $\phi_{\rm end}$ should be $N_{\rm CMB} \sim 50$:
\beq
N_{\rm CMB} &=& -\int_{\phi}^{\phi_{\rm end}} d\phi \frac{U (\phi)}{U'(\phi)} 
\\
&\simeq& \frac{N_{\rm max}}{\pi} {\rm Arctan} \lkk \frac{\sqrt{\rho/2\eta}(\phi_{\rm end}- \phi )}{1 + (\rho/2\eta) \phi \phi_{\rm end}} \rkk, 
\label{NO}
\eeq
where we assume $U(\phi) \simeq U$ in the denominator. 
Here
the maximal number of e-folds $N_{\rm max}$ is given by 
\beq
N_{\rm max} = - \int_{-\phi_{\rm end}}^{\phi_{\rm end}} d\phi \frac{U(\phi)}{U'(\phi)} \approx \pi\sqrt{2}\frac{U}{\sqrt{\eta\rho}}.
\label{Nmax2}
\eeq
The field value $\phi$ at which the CMB scale leaves the horizon is therefore 
given by 
\beq
 \sqrt{\frac{\rho}{2 \eta}} \phi = -\frac{\tan x + y}{1 - y \tan x}, 
\eeq
where 
\beq
 x \equiv \pi \frac{N_{\rm CMB}}{N_{\rm max}} 
 \label{x}
 \\
 y \equiv \frac{N_{\rm max}}{2 \pi}. 
 \label{y}
\eeq
The spectral index is given by 
\beq
 1 - n_s  &=& 6 \cdot \frac{1}{2} \lmk \frac{U' (\phi)}{U} \rmk^2 - 2 \lmk \frac{U''(\phi)}{U} \rmk 
 \\
 &\simeq& -\frac{2\rho \phi}{U} 
 = 
 \frac{2}{y} \frac{y + \tan x}{y \tan x  - 1}, 
 \label{n_s1}
\eeq
where we neglect the first term in the first line.

The magnitude of density fluctuations produced at $\phi$ is 
\beq
Q^2&=&\frac{1}{12\pi} \frac{U^3(\phi)}{{U'}^2 (\phi)} 
\\
&\simeq&
\frac{U^3}{12 \pi \eta^2} \frac{\lmk 1 - y \tan x \rmk^4}{\lkk \lmk 1 - y \tan x \rmk^2 + \lmk \tan x + y \rmk^2 \rkk^2}. 
\eeq
This can be rewritten as 
\beq
 &&Q^2 \simeq \frac{N_{\rm CMB}^4 \rho^2}{48 \pi U} f^2(x, y ) 
 \\
 &&f(x, y ) \equiv 
 \frac{\cos^2 x \lmk y \tan x  - 1\rmk^2 }{x^2 \lmk y^2 + 1 \rmk}. 
 \label{fxy1}
\eeq
Note that $x$ and $y$ depend only on $N_{\rm max}$. 
Note also that $f(x,y) \simeq 1$ for $N_{\rm max} \gtrsim N_{\rm CMB} \gg 1$. 
Also, $f(x,y) \sim 1$ for $y \gg 1$ and $x \sim 1$.

\subsection{Saddle point inflation} 

The potential for saddle point inflation is
\beq
U(\phi)=U + \frac{1}{2} \lambda \phi^2 + \frac{1}{3!}\rho \phi^3 +..., 
\eeq
where $\lambda <0$. 
The slow roll ends when $|U''|/U = 1$, or
\beq
\phi_{\rm end} = -\frac{ U + \lambda}{ \rho}.
\eeq

The number of e-folds from the observable scale $\phi$ to $\phi_{\rm end}$ should be $N_{\rm CMB} \sim 50$:
\beq
N_{\rm CMB} 
\simeq 
 \frac{U}{\abs{\lambda}} \ln \lmk \frac{\phi}{\phi_{\rm end}} \frac{1 - \rho \phi_{\rm end} / 2 \abs{\lambda}}{1 - \rho \phi / 2 \abs{\lambda}} \rmk. 
\eeq
Here
the ``maximal number of e-folds'' $N_{\rm max}$ is defined by 
\beq
N_{\rm max} \equiv 2 \pi \frac{U}{\abs{\lambda}}.
\eeq
The field value $\phi$ at which the CMB scale leaves the horizon is therefore 
given by 
\beq
 \frac{\rho}{2 \abs{\lambda}} \phi \simeq - \frac{(y-1) e^x}{(y+1) e^{-x} - (y-1) e^x}, 
\eeq
where we use $\rho \phi_{\rm end} / 2 \abs{\lambda} = - (y-1)/2$, 
and $x$ and $y$ are defined by Eqs.~(\ref{x}) and (\ref{y}), respectively.

The spectral index is given by 
\beq
 1 - n_s  &=& 6 \cdot \frac{1}{2} \lmk \frac{U' (\phi)}{U} \rmk^2 - 2 \lmk \frac{U''(\phi)}{U} \rmk 
 \\
 &\simeq& - 2 \frac{\lambda + \rho \phi}{U} 
 = 
 \frac{2}{y} \frac{y - \tanh x}{y \tanh x  - 1}, 
 \label{n_s2}
\eeq
where we neglect the first term in the first line.

The magnitude of density fluctuations produced at $\phi$ is 
\beq
Q^2
&\simeq&
\frac{U^3}{12 \pi \lambda^2} \frac{\rho^2}{4 \lambda^2} 
\frac{\lkk (y+1) e^{-x} - (y-1) e^x\rkk^4}{\lkk (y-1) e^x \rkk^2 \lkk (y+1) e^{-x} \rkk^2}, 
\eeq
This can be rewritten as 
\beq
 &&Q^2 \simeq \frac{N_{\rm CMB}^4 \rho^2}{48 \pi U}  f^2(x,y) 
 \label{Q2}
 \\
 &&f(x,y) \equiv 
 \frac{\cosh^2 x \lmk y \tanh x  - 1\rmk^2 }{x^2 \lmk y^2 - 1 \rmk}. 
 \label{fxy2}
\eeq
Note again that $x$ and $y$ depend only on $N_{\rm max}$ 
and $f(x,y) \to 1$ for $N_{\rm max} \gtrsim N_{\rm CMB} \gg 1$.

\section{Calculation of ${\cal P}_2' (\eta, \rho)$}
\label{sec:quadratic}

In this Appendix we calculate ${\cal P}_2' (\eta, \rho)$ 
in a large $D$ limit. 
Explicitly, it is given by 
\beq
 {\cal P}_2' (\eta, \rho)
 =
A_2 
\int 
\prod_{(ijk) \ne (111)} d \rho_{ijk}
\exp \lkk - c_5 \frac{\Lambda^2}{U_0^2} \eta \eta - c_6 \frac{\Lambda^4}{DU_0^2} \eta \rho_{1ii}  + 
c_7 \frac{\Lambda^6}{DU_0^2} \rho_{iik}\rho_{jjk} - c_8 \frac{\Lambda^6}{U_0^2} \rho_{ijk}\rho_{ijk}
\rkk~, \nonumber 
\eeq
where we have performed the integrals of the delta functions. 
Since we are interested in the distribution of $\eta$ and $\rho$, 
the integrals of $\rho_{ijk}$ with $i < j < k$, 
$1 < i  \ne  j = k$, or their permutations can be absorbed into the normalization factor. 
Then we obtain 
\beq
 {\cal P}_2' (\eta, \rho)
 =
A_2' 
\int 
\prod_{a \ge 2} d \rho_{1aa}
\exp \lkk - c_5 \frac{\Lambda^2}{U_0^2} \eta \eta - c_6 \frac{\Lambda^4}{DU_0^2} \eta \rho_{1ii}  + 
c_7 \frac{\Lambda^6}{DU_0^2} \rho_{1ii}\rho_{1jj} - 3c_8 \frac{\Lambda^6}{U_0^2} \sum_i \rho_{1ii}\rho_{1ii}
\rkk~, \nonumber 
\eeq
This is Gaussian integrals of $D-1$ variables $\rho_{1aa}$ ($a \ge 2$). 
The result is given by 
\beq 
 {\cal P}_2' (\eta, \rho)
 =
A_2'' 
\exp \lkk - c_5 \frac{\Lambda^2}{U_0^2} \eta^2 
- c_6 \frac{\Lambda^4}{DU_0^2} \eta \rho 
- (3 c_8 - c_7/ D)  \frac{\Lambda^6}{U_0^2} \rho^2 
+ C(\eta, \rho) 
\rkk, 
\eeq
where $C(\eta, \rho)$ comes from the integrals of $\rho_{1aa}$: 
\beq
 C(\eta, \rho) = 
  \lmk \frac{1}{D(3 c_8D  - c_7(D-1))} \rmk 
 \lmk 
 \frac{c_6^2}{4} \frac{\Lambda^2}{U_0^2} \eta^2 
 - c_6 c_7 \frac{\Lambda^2}{U_0^4} \eta \rho 
 + c_7^2 \frac{\Lambda^2}{U_0^6} \rho^2 
 \rmk. 
\eeq
Since $C(\eta, \rho)$ is suppressed by $1/D^2$, we can neglect it
in a large $D$ limit. The normalization constant can be easily determined by 
the dimensional analysis. 
Note first that ${\cal P}_2 (\eta_i, \rho_{ijk})$ is 
normalized as \eq{P_2 normalization}. 
On the other hand, $\eta$ and $\rho$ are not integrated out in ${\cal P}_2'$ 
and there are $\prod_{a \ge 2} \delta (\eta_i)$ 
in \eq{P_2p} for ${\cal P}_2'(\eta, \rho)$. 
So there should be 
a factor of $(U_0 / \Lambda)^{-1} * (U_0 / \Lambda^3)^{-1} * (U_0 / \Lambda)^{-(D-1)}$. 
Thus we estimate 
\beq
 A_2'' \sim \frac{\Lambda^{D +3}}{U_0^{D+1}}. 
\eeq

\bibliography{reference}

\providecommand{\href}[2]{#2}\begingroup\raggedright\begin{thebibliography}{10}

\bibitem{Douglas:2006es}
M.~R. Douglas and S.~Kachru, \emph{{Flux compactification}},
  \href{https://doi.org/10.1103/RevModPhys.79.733}{\emph{Rev. Mod. Phys.}
  {\bfseries 79} (2007) 733}
  [\href{https://arxiv.org/abs/hep-th/0610102}{{\ttfamily hep-th/0610102}}].

\bibitem{Bousso:2000xa}
R.~Bousso and J.~Polchinski, \emph{{Quantization of four form fluxes and
  dynamical neutralization of the cosmological constant}},
  \href{https://doi.org/10.1088/1126-6708/2000/06/006}{\emph{JHEP} {\bfseries
  06} (2000) 006} [\href{https://arxiv.org/abs/hep-th/0004134}{{\ttfamily
  hep-th/0004134}}].

\bibitem{Tegmark:2004qd}
M.~Tegmark, \emph{{What does inflation really predict?}},
  \href{https://doi.org/10.1088/1475-7516/2005/04/001}{\emph{JCAP} {\bfseries
  0504} (2005) 001} [\href{https://arxiv.org/abs/astro-ph/0410281}{{\ttfamily
  astro-ph/0410281}}].

\bibitem{Aazami:2005jf}
A.~Aazami and R.~Easther, \emph{{Cosmology from random multifield potentials}},
  \href{https://doi.org/10.1088/1475-7516/2006/03/013}{\emph{JCAP} {\bfseries
  0603} (2006) 013} [\href{https://arxiv.org/abs/hep-th/0512050}{{\ttfamily
  hep-th/0512050}}].

\bibitem{Frazer:2011tg}
J.~Frazer and A.~R. Liddle, \emph{{Exploring a string-like landscape}},
  \href{https://doi.org/10.1088/1475-7516/2011/02/026}{\emph{JCAP} {\bfseries
  1102} (2011) 026} [\href{https://arxiv.org/abs/1101.1619}{{\ttfamily
  1101.1619}}].

\bibitem{Battefeld:2012qx}
D.~Battefeld, T.~Battefeld and S.~Schulz, \emph{{On the Unlikeliness of
  Multi-Field Inflation: Bounded Random Potentials and our Vacuum}},
  \href{https://doi.org/10.1088/1475-7516/2012/06/034}{\emph{JCAP} {\bfseries
  1206} (2012) 034} [\href{https://arxiv.org/abs/1203.3941}{{\ttfamily
  1203.3941}}].

\bibitem{Yang:2012jf}
I.-S. Yang, \emph{{Probability of Slowroll Inflation in the Multiverse}},
  \href{https://doi.org/10.1103/PhysRevD.86.103537}{\emph{Phys. Rev.}
  {\bfseries D86} (2012) 103537}
  [\href{https://arxiv.org/abs/1208.3821}{{\ttfamily 1208.3821}}].

\bibitem{Bachlechner:2014rqa}
T.~C. Bachlechner, \emph{{On Gaussian Random Supergravity}},
  \href{https://doi.org/10.1007/JHEP04(2014)054}{\emph{JHEP} {\bfseries 04}
  (2014) 054} [\href{https://arxiv.org/abs/1401.6187}{{\ttfamily 1401.6187}}].

\bibitem{Wang:2015rel}
G.~Wang and T.~Battefeld, \emph{{Vacuum Selection on Axionic Landscapes}},
  \href{https://doi.org/10.1088/1475-7516/2016/04/025}{\emph{JCAP} {\bfseries
  1604} (2016) 025} [\href{https://arxiv.org/abs/1512.04224}{{\ttfamily
  1512.04224}}].

\bibitem{Masoumi:2016eqo}
A.~Masoumi and A.~Vilenkin, \emph{{Vacuum statistics and stability in axionic
  landscapes}},
  \href{https://doi.org/10.1088/1475-7516/2016/03/054}{\emph{JCAP} {\bfseries
  1603} (2016) 054} [\href{https://arxiv.org/abs/1601.01662}{{\ttfamily
  1601.01662}}].

\bibitem{Easther:2016ire}
R.~Easther, A.~H. Guth and A.~Masoumi, \emph{{Counting Vacua in Random
  Landscapes}},  \href{https://arxiv.org/abs/1612.05224}{{\ttfamily
  1612.05224}}.

\bibitem{Masoumi:2016eag}
A.~Masoumi, A.~Vilenkin and M.~Yamada, \emph{{Inflation in random Gaussian
  landscapes}},
  \href{https://doi.org/10.1088/1475-7516/2017/05/053}{\emph{JCAP} {\bfseries
  1705} (2017) 053} [\href{https://arxiv.org/abs/1612.03960}{{\ttfamily
  1612.03960}}].

\bibitem{Masoumi:2017gmh}
A.~Masoumi, A.~Vilenkin and M.~Yamada, \emph{{Initial conditions for slow-roll
  inflation in a random Gaussian landscape}},
  \href{https://doi.org/10.1088/1475-7516/2017/07/003}{\emph{JCAP} {\bfseries
  1707} (2017) 003} [\href{https://arxiv.org/abs/1704.06994}{{\ttfamily
  1704.06994}}].

\bibitem{Masoumi:2017xbe}
A.~Masoumi, A.~Vilenkin and M.~Yamada, \emph{{Inflation in multi-field random
  Gaussian landscapes}},
  \href{https://doi.org/10.1088/1475-7516/2017/12/035}{\emph{JCAP} {\bfseries
  1712} (2017) 035} [\href{https://arxiv.org/abs/1707.03520}{{\ttfamily
  1707.03520}}].

\bibitem{Bjorkmo:2017nzd}
T.~Bjorkmo and M.~C.~D. Marsh, \emph{{Manyfield Inflation in Random
  Potentials}},
  \href{https://doi.org/10.1088/1475-7516/2018/02/037}{\emph{JCAP} {\bfseries
  1802} (2018) 037} [\href{https://arxiv.org/abs/1709.10076}{{\ttfamily
  1709.10076}}].

\bibitem{Marsh:2013qca}
M.~C.~D. Marsh, L.~McAllister, E.~Pajer and T.~Wrase, \emph{{Charting an
  Inflationary Landscape with Random Matrix Theory}},
  \href{https://doi.org/10.1088/1475-7516/2013/11/040}{\emph{JCAP} {\bfseries
  1311} (2013) 040} [\href{https://arxiv.org/abs/1307.3559}{{\ttfamily
  1307.3559}}].

\bibitem{Dias:2016slx}
M.~Dias, J.~Frazer and M.~C.~D. Marsh, \emph{{Simple emergent power spectra
  from complex inflationary physics}},
  \href{https://doi.org/10.1103/PhysRevLett.117.141303}{\emph{Phys. Rev. Lett.}
  {\bfseries 117} (2016) 141303}
  [\href{https://arxiv.org/abs/1604.05970}{{\ttfamily 1604.05970}}].

\bibitem{Wang:2016kzp}
G.~Wang and T.~Battefeld, \emph{{Random Functions via Dyson Brownian Motion:
  Progress and Problems}},
  \href{https://doi.org/10.1088/1475-7516/2016/09/008}{\emph{JCAP} {\bfseries
  1609} (2016) 008} [\href{https://arxiv.org/abs/1607.02514}{{\ttfamily
  1607.02514}}].

\bibitem{Freivogel:2016kxc}
B.~Freivogel, R.~Gobbetti, E.~Pajer and I.-S. Yang, \emph{{Inflation on a
  Slippery Slope}},  \href{https://arxiv.org/abs/1608.00041}{{\ttfamily
  1608.00041}}.

\bibitem{Pedro:2016sli}
F.~G. Pedro and A.~Westphal, \emph{{Inflation with a graceful exit in a random
  landscape}}, \href{https://doi.org/10.1007/JHEP03(2017)163}{\emph{JHEP}
  {\bfseries 03} (2017) 163}
  [\href{https://arxiv.org/abs/1611.07059}{{\ttfamily 1611.07059}}].

\bibitem{Dias:2017gva}
M.~Dias, J.~Frazer and M.~c.~D. Marsh, \emph{{Seven Lessons from Manyfield
  Inflation in Random Potentials}},
  \href{https://doi.org/10.1088/1475-7516/2018/01/036}{\emph{JCAP} {\bfseries
  1801} (2018) 036} [\href{https://arxiv.org/abs/1706.03774}{{\ttfamily
  1706.03774}}].

\bibitem{BlancoPillado:2012cb}
J.~J. Blanco-Pillado, M.~Gomez-Reino and K.~Metallinos, \emph{{Accidental
  Inflation in the Landscape}},
  \href{https://doi.org/10.1088/1475-7516/2013/02/034}{\emph{JCAP} {\bfseries
  1302} (2013) 034} [\href{https://arxiv.org/abs/1209.0796}{{\ttfamily
  1209.0796}}].

\bibitem{M}
K.~Metallinos, \emph{{Numerical exploration of the string theory landscape}},
  {\emph{Phd Thesis, ProQuest Dissertations and Theses} {\bfseries 75-02(E)}
  (2013) 114}.

\bibitem{MartinezPedrera:2012rs}
D.~Martinez-Pedrera, D.~Mehta, M.~Rummel and A.~Westphal, \emph{{Finding all
  flux vacua in an explicit example}},
  \href{https://doi.org/10.1007/JHEP06(2013)110}{\emph{JHEP} {\bfseries 06}
  (2013) 110} [\href{https://arxiv.org/abs/1212.4530}{{\ttfamily 1212.4530}}].

\bibitem{Conlon:2005ki}
J.~P. Conlon, F.~Quevedo and K.~Suruliz, \emph{{Large-volume flux
  compactifications: Moduli spectrum and D3/D7 soft supersymmetry breaking}},
  \href{https://doi.org/10.1088/1126-6708/2005/08/007}{\emph{JHEP} {\bfseries
  08} (2005) 007} [\href{https://arxiv.org/abs/hep-th/0505076}{{\ttfamily
  hep-th/0505076}}].

\bibitem{Gallego:2008qi}
D.~Gallego and M.~Serone, \emph{{An Effective Description of the Landscape.
  I.}}, \href{https://doi.org/10.1088/1126-6708/2009/01/056}{\emph{JHEP}
  {\bfseries 01} (2009) 056} [\href{https://arxiv.org/abs/0812.0369}{{\ttfamily
  0812.0369}}].

\bibitem{Bray:2007tf}
A.~J. Bray and D.~S. Dean, \emph{{Statistics of critical points of Gaussian
  fields on large-dimensional spaces}},
  \href{https://doi.org/10.1103/PhysRevLett.98.150201}{\emph{Phys. Rev. Lett.}
  {\bfseries 98} (2007) 150201}.

\bibitem{Dvali:2001sm}
G.~R. Dvali and A.~Vilenkin, \emph{{Field theory models for variable
  cosmological constant}},
  \href{https://doi.org/10.1103/PhysRevD.64.063509}{\emph{Phys. Rev.}
  {\bfseries D64} (2001) 063509}
  [\href{https://arxiv.org/abs/hep-th/0102142}{{\ttfamily hep-th/0102142}}].

\bibitem{Dvali:2004tma}
G.~Dvali, \emph{{Large hierarchies from attractor vacua}},
  \href{https://doi.org/10.1103/PhysRevD.74.025018}{\emph{Phys. Rev.}
  {\bfseries D74} (2006) 025018}
  [\href{https://arxiv.org/abs/hep-th/0410286}{{\ttfamily hep-th/0410286}}].

\bibitem{Brodie:2015kza}
C.~Brodie and M.~C.~D. Marsh, \emph{{The Spectra of Type IIB Flux
  Compactifications at Large Complex Structure}},
  \href{https://doi.org/10.1007/JHEP01(2016)037}{\emph{JHEP} {\bfseries 01}
  (2016) 037} [\href{https://arxiv.org/abs/1509.06761}{{\ttfamily
  1509.06761}}].

\bibitem{Marsh:2015zoa}
M.~C.~D. Marsh and K.~Sousa, \emph{{Universal Properties of Type IIB and
  F-theory Flux Compactifications at Large Complex Structure}},
  \href{https://doi.org/10.1007/JHEP03(2016)064}{\emph{JHEP} {\bfseries 03}
  (2016) 064} [\href{https://arxiv.org/abs/1512.08549}{{\ttfamily
  1512.08549}}].

\bibitem{Dine:1985he}
M.~Dine and N.~Seiberg, \emph{{Is the Superstring Weakly Coupled?}},
  \href{https://doi.org/10.1016/0370-2693(85)90927-X}{\emph{Phys. Lett.}
  {\bfseries 162B} (1985) 299}.

\bibitem{Bachlechner:2016mtp}
T.~C. Bachlechner, \emph{{Inflation Expels Runaways}},
  \href{https://doi.org/10.1007/JHEP12(2016)155}{\emph{JHEP} {\bfseries 12}
  (2016) 155} [\href{https://arxiv.org/abs/1608.07576}{{\ttfamily
  1608.07576}}].

\bibitem{Freivogel:2011eg}
B.~Freivogel, \emph{{Making predictions in the multiverse}},
  \href{https://doi.org/10.1088/0264-9381/28/20/204007}{\emph{Class. Quant.
  Grav.} {\bfseries 28} (2011) 204007}
  [\href{https://arxiv.org/abs/1105.0244}{{\ttfamily 1105.0244}}].

\bibitem{Linde:1993nz}
A.~D. Linde and A.~Mezhlumian, \emph{{Stationary universe}},
  \href{https://doi.org/10.1016/0370-2693(93)90187-M}{\emph{Phys. Lett.}
  {\bfseries B307} (1993) 25}
  [\href{https://arxiv.org/abs/gr-qc/9304015}{{\ttfamily gr-qc/9304015}}].

\bibitem{Linde:1993xx}
A.~D. Linde, D.~A. Linde and A.~Mezhlumian, \emph{{From the Big Bang theory to
  the theory of a stationary universe}},
  \href{https://doi.org/10.1103/PhysRevD.49.1783}{\emph{Phys. Rev.} {\bfseries
  D49} (1994) 1783} [\href{https://arxiv.org/abs/gr-qc/9306035}{{\ttfamily
  gr-qc/9306035}}].

\bibitem{DeSimone:2008bq}
A.~De~Simone, A.~H. Guth, M.~P. Salem and A.~Vilenkin, \emph{{Predicting the
  cosmological constant with the scale-factor cutoff measure}},
  \href{https://doi.org/10.1103/PhysRevD.78.063520}{\emph{Phys. Rev.}
  {\bfseries D78} (2008) 063520}
  [\href{https://arxiv.org/abs/0805.2173}{{\ttfamily 0805.2173}}].

\bibitem{Bousso:2008hz}
R.~Bousso, B.~Freivogel and I.-S. Yang, \emph{{Properties of the scale factor
  measure}}, \href{https://doi.org/10.1103/PhysRevD.79.063513}{\emph{Phys.
  Rev.} {\bfseries D79} (2009) 063513}
  [\href{https://arxiv.org/abs/0808.3770}{{\ttfamily 0808.3770}}].

\bibitem{Garriga:2012bc}
J.~Garriga and A.~Vilenkin, \emph{{Watchers of the multiverse}},
  \href{https://doi.org/10.1088/1475-7516/2013/05/037}{\emph{JCAP} {\bfseries
  1305} (2013) 037} [\href{https://arxiv.org/abs/1210.7540}{{\ttfamily
  1210.7540}}].

\bibitem{Vilenkin:2013loa}
A.~Vilenkin, \emph{{A quantum measure of the multiverse}},
  \href{https://doi.org/10.1088/1475-7516/2014/05/005}{\emph{JCAP} {\bfseries
  1405} (2014) 005} [\href{https://arxiv.org/abs/1312.0682}{{\ttfamily
  1312.0682}}].

\bibitem{DeSimone:2008if}
A.~De~Simone, A.~H. Guth, A.~D. Linde, M.~Noorbala, M.~P. Salem and
  A.~Vilenkin, \emph{{Boltzmann brains and the scale-factor cutoff measure of
  the multiverse}},
  \href{https://doi.org/10.1103/PhysRevD.82.063520}{\emph{Phys. Rev.}
  {\bfseries D82} (2010) 063520}
  [\href{https://arxiv.org/abs/0808.3778}{{\ttfamily 0808.3778}}].

\bibitem{Garriga:1997ef}
J.~Garriga and A.~Vilenkin, \emph{{Recycling universe}},
  \href{https://doi.org/10.1103/PhysRevD.57.2230}{\emph{Phys. Rev.} {\bfseries
  D57} (1998) 2230} [\href{https://arxiv.org/abs/astro-ph/9707292}{{\ttfamily
  astro-ph/9707292}}].

\bibitem{Garriga:2005av}
J.~Garriga, D.~Schwartz-Perlov, A.~Vilenkin and S.~Winitzki,
  \emph{{Probabilities in the inflationary multiverse}},
  \href{https://doi.org/10.1088/1475-7516/2006/01/017}{\emph{JCAP} {\bfseries
  0601} (2006) 017} [\href{https://arxiv.org/abs/hep-th/0509184}{{\ttfamily
  hep-th/0509184}}].

\bibitem{Garriga:2013cix}
J.~Garriga, A.~Vilenkin and J.~Zhang, \emph{{Non-singular bounce transitions in
  the multiverse}},
  \href{https://doi.org/10.1088/1475-7516/2013/11/055}{\emph{JCAP} {\bfseries
  1311} (2013) 055} [\href{https://arxiv.org/abs/1309.2847}{{\ttfamily
  1309.2847}}].

\bibitem{Masoumi:2016wot}
A.~Masoumi, K.~D. Olum and B.~Shlaer, \emph{{Efficient numerical solution to
  vacuum decay with many fields}},
  \href{https://doi.org/10.1088/1475-7516/2017/01/051}{\emph{JCAP} {\bfseries
  1701} (2017) 051} [\href{https://arxiv.org/abs/1610.06594}{{\ttfamily
  1610.06594}}].

\bibitem{Linde:2016uec}
A.~Linde, \emph{{Random Potentials and Cosmological Attractors}},
  \href{https://doi.org/10.1088/1475-7516/2017/02/028}{\emph{JCAP} {\bfseries
  1702} (2017) 028} [\href{https://arxiv.org/abs/1612.04505}{{\ttfamily
  1612.04505}}].

\bibitem{Weinberg:1987dv}
S.~Weinberg, \emph{{Anthropic Bound on the Cosmological Constant}},
  \href{https://doi.org/10.1103/PhysRevLett.59.2607}{\emph{Phys. Rev. Lett.}
  {\bfseries 59} (1987) 2607}.

\bibitem{Weinberg:1988cp}
S.~Weinberg, \emph{{The Cosmological Constant Problem}},
  \href{https://doi.org/10.1103/RevModPhys.61.1}{\emph{Rev. Mod. Phys.}
  {\bfseries 61} (1989) 1}.

\bibitem{Linde:2007jn}
A.~D. Linde and A.~Westphal, \emph{{Accidental Inflation in String Theory}},
  \href{https://doi.org/10.1088/1475-7516/2008/03/005}{\emph{JCAP} {\bfseries
  0803} (2008) 005} [\href{https://arxiv.org/abs/0712.1610}{{\ttfamily
  0712.1610}}].

\bibitem{Baumann:2007ah}
D.~Baumann, A.~Dymarsky, I.~R. Klebanov and L.~McAllister, \emph{{Towards an
  Explicit Model of D-brane Inflation}},
  \href{https://doi.org/10.1088/1475-7516/2008/01/024}{\emph{JCAP} {\bfseries
  0801} (2008) 024} [\href{https://arxiv.org/abs/0706.0360}{{\ttfamily
  0706.0360}}].

\bibitem{Vilenkin:1983xq}
A.~Vilenkin, \emph{{The Birth of Inflationary Universes}},
  \href{https://doi.org/10.1103/PhysRevD.27.2848}{\emph{Phys. Rev.} {\bfseries
  D27} (1983) 2848}.

\bibitem{Agarwal:2011wm}
N.~Agarwal, R.~Bean, L.~McAllister and G.~Xu, \emph{{Universality in D-brane
  Inflation}}, \href{https://doi.org/10.1088/1475-7516/2011/09/002}{\emph{JCAP}
  {\bfseries 1109} (2011) 002}
  [\href{https://arxiv.org/abs/1103.2775}{{\ttfamily 1103.2775}}].

\bibitem{Fyodorov}
Y.~V. Fyodorov, \emph{{Complexity of Random Energy Landscapes, Glass Transition
  and Absolute Value of Spectral Determinant of Random Matrices}}, {\emph{Phys.
  Rev. Lett.} {\bfseries 92} (2004) }.

\bibitem{Yamada:2017uzq}
M.~Yamada and A.~Vilenkin, \emph{{Hessian eigenvalue distribution in a random
  Gaussian landscape}},
  \href{https://doi.org/10.1007/JHEP03(2018)029}{\emph{JHEP} {\bfseries 03}
  (2018) 029} [\href{https://arxiv.org/abs/1712.01282}{{\ttfamily
  1712.01282}}].

\bibitem{Vilenkin:1996ar}
A.~Vilenkin and S.~Winitzki, \emph{{Probability distribution for omega in open
  universe inflation}},
  \href{https://doi.org/10.1103/PhysRevD.55.548}{\emph{Phys. Rev.} {\bfseries
  D55} (1997) 548} [\href{https://arxiv.org/abs/astro-ph/9605191}{{\ttfamily
  astro-ph/9605191}}].

\bibitem{Freivogel:2005vv}
B.~Freivogel, M.~Kleban, M.~Rodriguez~Martinez and L.~Susskind,
  \emph{{Observational consequences of a landscape}},
  \href{https://doi.org/10.1088/1126-6708/2006/03/039}{\emph{JHEP} {\bfseries
  03} (2006) 039} [\href{https://arxiv.org/abs/hep-th/0505232}{{\ttfamily
  hep-th/0505232}}].

\bibitem{Tegmark:1997in}
M.~Tegmark and M.~J. Rees, \emph{{Why is the Cosmic Microwave Background
  fluctuation level 10**(-5)?}},
  \href{https://doi.org/10.1086/305673}{\emph{Astrophys. J.} {\bfseries 499}
  (1998) 526} [\href{https://arxiv.org/abs/astro-ph/9709058}{{\ttfamily
  astro-ph/9709058}}].

\bibitem{Tegmark:2005dy}
M.~Tegmark, A.~Aguirre, M.~Rees and F.~Wilczek, \emph{{Dimensionless constants,
  cosmology and other dark matters}},
  \href{https://doi.org/10.1103/PhysRevD.73.023505}{\emph{Phys. Rev.}
  {\bfseries D73} (2006) 023505}
  [\href{https://arxiv.org/abs/astro-ph/0511774}{{\ttfamily
  astro-ph/0511774}}].

\bibitem{Martel:1997vi}
H.~Martel, P.~R. Shapiro and S.~Weinberg, \emph{{Likely values of the
  cosmological constant}},
  \href{https://doi.org/10.1086/305016}{\emph{Astrophys. J.} {\bfseries 492}
  (1998) 29} [\href{https://arxiv.org/abs/astro-ph/9701099}{{\ttfamily
  astro-ph/9701099}}].

\bibitem{Garriga:2002tq}
J.~Garriga and A.~Vilenkin, \emph{{Testable anthropic predictions for dark
  energy}}, \href{https://doi.org/10.1103/PhysRevD.67.043503}{\emph{Phys. Rev.}
  {\bfseries D67} (2003) 043503}
  [\href{https://arxiv.org/abs/astro-ph/0210358}{{\ttfamily
  astro-ph/0210358}}].

\bibitem{Garriga:1999hu}
J.~Garriga, M.~Livio and A.~Vilenkin, \emph{{The Cosmological constant and the
  time of its dominance}},
  \href{https://doi.org/10.1103/PhysRevD.61.023503}{\emph{Phys. Rev.}
  {\bfseries D61} (2000) 023503}
  [\href{https://arxiv.org/abs/astro-ph/9906210}{{\ttfamily
  astro-ph/9906210}}].

\bibitem{Piran:2014wfa}
T.~Piran and R.~Jimenez, \emph{{Possible Role of Gamma Ray Bursts on Life
  Extinction in the Universe}},
  \href{https://doi.org/10.1103/PhysRevLett.113.231102}{\emph{Phys. Rev. Lett.}
  {\bfseries 113} (2014) 231102}
  [\href{https://arxiv.org/abs/1409.2506}{{\ttfamily 1409.2506}}].

\bibitem{Bachlechner:2017hsj}
T.~C. Bachlechner, K.~Eckerle, O.~Janssen and M.~Kleban, \emph{{Systematics of
  Aligned Axions}}, \href{https://doi.org/10.1007/JHEP11(2017)036}{\emph{JHEP}
  {\bfseries 11} (2017) 036}
  [\href{https://arxiv.org/abs/1709.01080}{{\ttfamily 1709.01080}}].

\end{thebibliography}\endgroup

\end{document}